\documentclass[aps,prd, notitlepage, onecolumn,superscriptaddress,floatfix,letterpaper,nofootinbib, longbibliography]{revtex4-1}

\usepackage{amsmath, amsthm, amssymb,slashed}

\usepackage{comment}

\usepackage[usenames, dvipsnames]{color}
\usepackage[svgnames]{xcolor}
\usepackage[colorlinks,citecolor=RoyalBlue, urlcolor=RoyalBlue, linkcolor=RoyalBlue ]{hyperref} 

\usepackage[normalem]{ulem}

\newcommand{\cpurple}[1]{\textcolor{purple}{#1}}

\usepackage{booktabs}

\definecolor{mygray}{gray}{0.6}

\usepackage{upgreek}
\usepackage{bbm}

\newenvironment{myfont}[2][]{\csname#2\endcsname[#1]}{}

\usepackage{slashed}
\usepackage[makeroom]{cancel}
\usepackage[normalem]{ulem}
\usepackage{soul}
\newcommand{\stkout}[1]{\ifmmode\text{\sout{\ensuremath{#1}}}\else\sout{#1}\fi}

\usepackage{sseq}
\usepackage[all,cmtip]{xy}
\usepackage{tikz-cd}
\usepackage{tikz}
\usetikzlibrary{matrix}
\usetikzlibrary{decorations.markings}
\usetikzlibrary{tikzmark,decorations.pathreplacing,positioning}
\usepackage{amsfonts}
\usepackage{multirow}

\newcommand{\bea}{\begin{eqnarray}}
\newcommand{\eea}{\end{eqnarray}}
\def\be{\begin{equation}}
\def\ee{\end{equation}}

\newcommand{\e}{\hspace{1pt}\mathrm{e}}

\newcommand{\ii}{\hspace{1pt}\mathrm{i}\hspace{1pt}}

\definecolor{red}{rgb}{1,0,0}
\definecolor{blue}{rgb}{0,0,1}
\definecolor{dblue}{rgb}{0,0,0.4}
\definecolor{green}{rgb}{0,1,0}
\definecolor{black}{rgb}{0,0,0}
\definecolor{white}{rgb}{1,1,1}

\definecolor{brn}{rgb}{.8,.4,.0}
\definecolor{redo}{rgb}{1,.5,.0}
\definecolor{ddgrn}{rgb}{0,0.4,0}
\definecolor{dgrn}{rgb}{0,0.55,0}
\definecolor{dbl}{rgb}{0,0,0.5}

\usepackage[bbgreekl]{mathbbol}
\usepackage{amscd}

\newcommand{\Z}{\mathbb{Z}}
\newcommand{\C}{\mathbb{C}}
\newcommand{\R}{\mathbb{R}}

\newcommand{\dd}{\mathrm{d}}

\newcommand{\Eq}[1]{Eq.~(\ref{#1})} 
\newcommand{\eq}[1]{eq.~(\ref{#1})} 
\newcommand{\eqq}[1]{(\ref{#1})}

\newcommand{\Tr}{{\rm Tr}}

\newcommand{\bpm}{\begin{pmatrix}}
\newcommand{\epm}{\end{pmatrix}}
\newcommand{\bmm}{\begin{matrix}}
\newcommand{\emm}{\end{matrix}}

\newcommand{\cD}{ {\cal D} } 
 
\newcommand{\cG}{ {\cal G} }

\def\Z{{\mathbb{Z}}}

\def\R{{\mathbb{R}}}
\def\C{{\mathbb{C}}}

\def\Tr{{\mathrm{Tr}}}

\def \H{\operatorname{H}}

\def \Z{\mathbb{Z}}

\newcommand {\emptycomment}[1]{}

\def\TP{\mathrm{TP}}

\def\B{\mathrm{B}}

\newcommand{\SO}{{\rm SO}}
\newcommand{\Spin}{{\rm Spin}}
\newcommand{\U}{{\rm U}}
\newcommand{\SU}{{\rm SU}}

\usepackage{centernot}

\newcommand{\Sec}[1]{Sec.~\ref{#1}}

\newcommand{\diag}{{\rm diag}}
\usepackage{enumitem} 
\usepackage{mathtools,amssymb,varwidth}

\newcommand{\Fig}[1]{Fig.~\ref{#1}} 
\newcommand{\Table}[1]{Table \ref{#1}}

\usepackage{datetime}

\newcommand{\rF}{{\rm F}}

\usepackage{array}

\begin{document}


\title{Anomalous (3+1)d Fermionic
Topological 
Quantum Field Theories\\
via Symmetry Extension
}

\author{Zheyan Wan}
\email{wanzheyan@bimsa.cn}
\affiliation{Beijing Institute of Mathematical Sciences and Applications, Beijing 101408, China}

\author{Juven Wang}
\email{jw@lims.ac.uk}
\homepage{http://idear.info/}
\affiliation{London Institute for Mathematical Sciences, Royal Institution, W1S 4BS, UK}
\affiliation{Center of Mathematical Sciences and Applications, Harvard University, MA 02138, USA}


\begin{abstract}

Discrete finite-group global symmetries may suffer from nonperturbative ’t~Hooft anomalies. 
That nonperturbative global anomaly
may be potentially canceled by anomalous  symmetry-preserving topological quantum field theories (TQFTs), where the TQFTs contain no local 0d point operators (familiar to particle physics) but only extended operators such as 1d lines, 2d surfaces, etc., beyond particle physics.
In this work, we focus on the 
mixed gauge-gravitational 
nonperturbative global anomaly of the Weyl fermions 
(or Weyl semi-metal in condensed matter)
charged under a discrete abelian internal symmetry
in 4-dimensional spacetime, forming the spacetime-internal fermionic
symmetry 
$G =\mathrm{Spin} \times_{\mathbb{Z}_2^{\rm F}} \mathbb{Z}_{2m}^{\rm F}$ or
$\mathrm{Spin} \times \mathbb{Z}_n$
that contains fermion parity
$\mathbb{Z}_{2}^{\rm F}$.
We determine the minimal gauge group as a finite group $K$ of the anomalous $G$-symmetry-preserving TQFTs
that can match the Weyl fermion's anomaly
based on the symmetry-extension construction
$1 \to K \to G_{\rm Tot} \to G \to 1,$
via the trivialization of the $G$-anomaly index in 
the pullback $G_{\rm Tot}$.
{In practice, we apply 
the symmetry-extension
method to trivialize the nonperturbative global anomaly index computed by Atiyah-Patodi-Singer eta invariant 
in $G$ via pulling back to $G_{\rm Tot}$ by the group extension through $K$.}
Namely, for the anomaly matching
within the $G$-symmetry,
we can replace the $G$-symmetric 4d Weyl fermion by
anomalous $G$-symmetric discrete-$K$-gauge 4d TQFTs
as the alternative 
low-energy theories in the same deformation class.
As an example, the
4d Standard Model with 15 Weyl fermions per family and without 
the 16th Weyl fermion sterile right-handed neutrino $\nu_R$,
suffers from these types of nonperturbative global anomalies 
between the baryon plus or minus lepton number ${\bf B} \pm {\bf L}$ symmetry and gravitational diffeomorphism. 
Therefore, our work determines the appropriate or minimal $K$-gauge
fermionic TQFT as a low-energy theory of
topologically ordered 
quantum dark matter that replaces multiple 
Weyl fermions (such as $\nu_R$) of the SM
via the symmetry-extension construction, 
without involving the conventional Anderson-Higgs symmetry-breaking.

\end{abstract}


\maketitle

\tableofcontents



\newpage
\section{Introduction and Summary}

Topological Order \cite{Wen2016ddy1610.03911} is a 
long-range entangled (LRE)
gapped phase of many-body quantum matter whose low-energy theory can contain topological quantum field theories (TQFTs) \cite{Atiyah:1989vuTQFT}; in particular, non-invertible TQFTs,
whose mathematical description evokes often the category theory.

Symmetry-Protected Topological state (SPTs) \cite{Senthil1405.4015, Wen2016ddy1610.03911}
is a short-range entangled (SRE)
gapped phase of many-body quantum matter whose low-energy theory can contain, at most, invertible TQFTs.
We often call invertible TQFTs as invertible topological field theories (iTFTs), because
it can be described by an invertible partition function of a non-dynamical classical background field 
(instead of a dynamical quantum field).\footnote{The non-invertible TQFTs are those TQFTs whose partition function $Z(M)$ on a closed manifold $M$ has $|Z(M)| \neq 1$. 
The invertible TQFTs are those TQFTs whose partition function $Z(M)$ on a closed manifold $M$ has $|Z(M)| = 1$, thus $Z(M)=\e^{\ii \theta}$ just a complex phase in $\U(1)$ which in some sense, detects the anomaly when $\e^{\ii \theta} \neq 1$. }
The mathematical description of SPTs and iTFT evokes often the cohomology and cobordism theory.

The goal of this work aims to describe the appropriate $G$-symmetric anomalous LRE Topological Order living in the
(3+1)-dimensional spacetime [(3+1)d], whose 't Hooft anomaly of global $G$-symmetry \cite{tHooft1979ratanomaly}
can match with the
(3+1)d boundary state of the (4+1)d SRE bulk
Symmetry-Protected Topological state (SPTs).
Namely, we describe the appropriate $G$-symmetric anomalous \emph{non-invertible} TQFTs living in the
(3+1)d spacetime, whose 't Hooft anomaly of global $G$-symmetry can match with the
(3+1)d boundary state of a (4+1)d bulk \emph{invertible} TQFTs.
The generic phenomena of LRE boundary topological order and SRE bulk SPTs
is firstly explored in the (2+1)d LRE boundary and (3+1)d SRE bulk system in condensed matter physics suggested by
Vishwanath-Senthil \cite{VishwanathSenthil1209.3058},
and reviewed in \cite{Senthil1405.4015}.
Thus, we are exploring this analogous quantum condensed matter phenomena of LRE-boundary/SRE-bulk coupled system 
in one-higher-dimension as (3+1)d LRE boundary TQFT and (4+1)d SRE bulk iTFT system.
In this work, we are particularly interested in the case where the 't Hooft anomaly of 
the global symmetry $G$ in
(3+1)d has the following features:
\begin{enumerate}
\item {\bf  Fermionic anomaly } 
(instead of bosonic anomaly): In the sense that the fermion parity
$\mathbb{Z}_{2}^{\rm F}$ is a part of the global symmetry which is generated by $(-1)^{\rF}$ which acts as
$(-1)^{\rF}: f \to -  f$ on all fermionic point operators $f$ with a fermionic $(-1)$ sign,
while $(-1)^{\rF}$ leaves all bosonic operators $b \to + b$ untouched. Thus we consider 
those $G \supseteq \Z_2^\rF$.
\item {\bf Nonperturbative global anomaly}: We know that the perturbative local anomaly can be computed from a one-loop Feynman diagram coefficient, but the perturbative local anomaly \emph{cannot} be canceled by a symmetric gapped TQFT
living in the same dimension. 
In contrast, a nonperturbative global anomaly (e.g., \cite{Witten1985xe, 
WangWenWitten2018qoy1810.00844, Witten2019bou1909.08775}) 
may be potentially 
be canceled by a symmetric gapped TQFT
living in the same dimension, with certain criteria like the existence of symmetry-extension trivialization \cite{Witten2016cio1605.02391, Wang2017locWWW1705.06728}
(that we will explain in depth later) 
or other topology constraints \cite{Cordova2019bsd1910.04962, Cordova1912.13069}. 
For example, for fermions with an U(1) internal global symmetry,
and with the spacetime Lorentz symmetry Spin group,\footnote{Note that the
Spin as the Spin group has $\Spin = \Spin(d,1)$ for Lorentzian signature and
$\Spin = \Spin(d+1)$ for Euclidean signature, in $(d+1)$ spacetime dimensions.
We will make the choice of Lorentzian or Euclidean implicit, so that the readers 
can choose the signatures based on the context.}
the possible combined
global symmetry can be:
\bea \label{eq:SpinU1}
G =\mathrm{Spin} \times_{\mathbb{Z}_2^{\rm F}} \U(1)  \equiv \Spin^c
\text{ or } 
G= \mathrm{Spin} \times \U(1).
\eea
But the 't Hooft anomalies for fermions 
in (3+1)d\footnote{These (3+1)d fermions in Lorentzian signature 
include Weyl fermions ($2^{\mathbb{C}}_L$), Majorana fermions ($4^{\mathbb{R}}$), or Dirac fermions ($4^{\mathbb{C}}=2^{\mathbb{C}}_L \oplus 2^{\mathbb{C}}_R$) (traditionally \cite{Polchinski:1998rrString2} and see modern perspectives \cite{Wan:2023nqe2312.17126, Li:2024jle2412.11958, Li:2024dpq2412.19691}). Since the Weyl fermion is the (minimal) irreducible spinor representation (which has the same degrees of freedom as a Majorana fermion 
($2^{\mathbb{C}}_L \sim 4^{\mathbb{R}}$, and which only has $1/2$ degrees of freedom of a Dirac fermion),
we will focus on the anomalies of (multiple of) Weyl fermions.} 
of these symmetries \eqq{eq:SpinU1} that contain U(1) cannot be
nonperturbative global anomalies, but only perturbative local anomalies.
Instead, we consider the abelian cyclic subgroup of U(1), which is a finite discrete abelian group $\Z_n \equiv \Z/(n \Z)
\subset \U(1)$, so \eq{eq:SpinU1} becomes
\bea \label{eq:SpinZn}
G =\mathrm{Spin} \times_{\mathbb{Z}_2^{\rm F}} \mathbb{Z}_{2m}^{\rm F} 
\equiv
\frac{\Spin \times \mathbb{Z}_{2m}^{\rm F}}{{\mathbb{Z}_2^{\rm F}} }
\equiv 
\Spin\text{-}\mathbb{Z}_{2m}^{\rm F}
\text{ or } G= \mathrm{Spin} \times \mathbb{Z}_n.
\eea
\end{enumerate}
As we desire, happily, now 't Hooft anomalies  
in (3+1)d
of these symmetries \eqq{eq:SpinZn} cannot be
 perturbative local anomalies, but can only be nonperturbative global anomalies. In this work,                we will then apply the appropriate 
 symmetry-extension trivialization method \cite{Wang2017locWWW1705.06728},
 making 
a nonperturbative global anomaly 
 in $G$
 becomes anomaly-free in an appropriate
 $G_{\rm Tot}$ 
 via an appropriate group extension
\bea \label{eq:}
1 \to K \to G_{\rm Tot} \overset{r}{\longrightarrow} 
G 
\to
1.
\eea
Namely, the precise mathematical question is, 
given a $G$, we search for
what the appropriate $K$ and $G_{\rm Tot}$ are,
such that
a nonperturbative global anomaly index 
\bea
\nu_G \in \TP_d(G)
\eea 
in the Freed-Hopkins version
\cite{1604.06527} of cobordism group TP
(see \Sec{sec:Notations})
becomes the trivial anomaly class 
\bea
(r^*\nu)_{G_{\rm Tot}} =0 \in \TP_d(G_{\rm Tot})
\eea
for the cobordism group TP of the pulled back 
$G_{\rm Tot}$.
Here $r$ is the reduction map from $G_{\rm Tot} \overset{r}{\longrightarrow} 
G$,
then the $r^*$ with a $*$ denote the 
pullback.
According to  
\cite{Wang2017locWWW1705.06728}, this provides 
a 
(3+1)d anomalous $G$-symmetric  
 $K$-gauge topological order construction
 whose low-energy theory is a (3+1)d $K$-gauge TQFT, which is designed to carry the original
 't Hooft anomaly index
  in $G$, namely $\nu_G \in \TP_d(G)$.
                                                                                                                                                                                                                                                                                                                                                                                                                                                                                                                                                                                                                                                                                                                                                                                                                                                                                                                                                                                                                                                                                                                                                                                                                                                                                                                                                                                                                                                                                                                                                                                                                                                                                                     
What motivates us to study this phenomena of 
(3+1)d LRE boundary and (4+1)d SRE bulk system 
is two folds. 
The {\bf first motivation} is that our (3+1)d LRE TQFT state is an alterative boundary state of some 
(3+1)d Weyl fermion or Weyl semi-metal with discrete symmetries, or the quantum phase that
(3+1)d Weyl fermion/semi-metal can deform into after quantum phase transition under the same deformation class of quantum field theory that preserves the same 't Hooft anomaly.
The {\bf second motivation}, in addition to presenting formal curiosity-driven models or arbitrary set of Weyl fermion/semi-metal in condensed matter, 
we have found direct applications of our results to the Standard Model (SM) of particle physics.
Because of the following two reasons:
\begin{enumerate}
\item SM lacks of the three hypothetical 
right-handed neutrinos (${\nu_{e,R}}, {\nu_{\mu,R}}, {\nu_{\tau,R}}$).

SM with 15 Weyl fermions per family and without 
the 16th Weyl fermion sterile right-handed neutrino $\nu_R$ (for $e,\mu,\tau$ flavors),
suffers from the perturbative local
mixed-gauge-gravitational anomalies 
\cite{Eguchi:1976db, AlvarezGaume1983igWitten1984,
Alvarez-Gaume:1984zlq}
between the lepton number ${\bf L}$ symmetry and gravitational background fields in 3+1d spacetime. 
Namely these anomalies are computable
via perturbative triangle Feynman diagrams
with $\U(1)_{\bf L}^3$ and $\U(1)_{\bf L}$-gravity$^2$ on vertices, with the anomaly index coefficient 
\bea
-N_f + n_{\nu_R} = -3 + n_{\nu_R},
\eea
counting the difference between the family or generation number $N_f$ (typically $N_f=3$) and the total right-hand neutrino number $n_{\nu_R}$. See recent related expositions about this anomaly index 
$-N_f + n_{\nu_R}$
in  \cite{JW2006.16996, JW2008.06499, JW2012.15860, WangWanYou2112.14765, WangWanYou2204.08393, Putrov:2023jqi2302.14862, Wang:2024auy2501.00607}.

\item SM has various compatible Baryon minus Lepton (${\bf B} - {\bf L}$) or discrete Baryon plus Lepton 
(${\bf B} + {\bf L}$) symmetries:
\begin{enumerate}
\item 
For the ${\bf B} - {\bf L}$ symmetry, 
there is a full 
faithful $\U(1)_{{\bf B} - {\bf L}}$
symmetry for the gauge-invariant baryons,
or an unfaithful $\U(1)_{{\bf Q} - N_c {\bf L}}$
symmetry for the gauge-invariant baryons.
But $\U(1)_{{\bf Q} - N_c {\bf L}}$ is faithful for the free quarks,
where their charges are related by the $N_c$ color as
${\bf Q} - N_c {\bf L}=N_c({{\bf B} - {\bf L}})$),
 survived under the SM electroweak gauge instanton, see Table \ref{table:SM}.

\item
For the ${\bf B} + {\bf L}$ symmetry, 
there is a full 
faithful $\Z_{2 N_f, {\bf B} + {\bf L}}$
symmetry 
\cite{KorenProtonStability2204.01741,
Koren:2022axd2204.01750,
WangWanYou2204.08393}
for the gauge-invariant baryons
or an unfaithful $\Z_{2 N_c N_f, {\bf Q} + N_c {\bf L}}$
symmetry for the gauge-invariant baryons
(but $\Z_{2 N_c N_f, {\bf Q} + N_c {\bf L}}$ is faithful for the free quarks,
where their charges are related by the $N_c$ color as
${\bf Q} + N_c {\bf L}=N_c({{\bf B} + {\bf L}})$),
 survived under the SM electroweak gauge instanton,
see Table \ref{table:SM}.
\end{enumerate}
The combined symmetry of ${\bf B} - {\bf L}$ and ${\bf B} + {\bf L}$
is a full 
faithful 
\bea
\U(1)_{{\bf B} - {\bf L}} \times_{\Z_2^\rF}\Z_{2 N_f, {\bf B} + {\bf L}}
\eea
for the gauge-invariant baryons.
Again, $\times_{\Z_2^\rF}$
means a product group but mod out 
the fermion parity ${\Z_2^\rF}$.

The combined symmetry of ${\bf Q} - N_c {\bf L}$ and ${\bf Q} + N_c {\bf L}$
is a 
faithful 
\bea
\U(1)_{{\bf Q} - N_c {\bf L}} \times_{\Z_2^\rF}\Z_{2 N_c N_f, {\bf Q} + N_c {\bf L}}
\eea
for the free quarks, but it is unfaithful for the gauge-invariant baryons. 
Follow \cite{Wang:2025oow2502.21319}, 
  the $\U(1)$ symmetry can be broken down by the multi-fermion BSM interaction terms. So say
  $\U(1)_{{\bf B} - {\bf L}}$
  can be broken down to
  $\Z_{2m,{\bf B} - {\bf L}}$,
  or
  $\U(1)_{{\bf Q} - N_c {\bf L}}$
  can be broken down to
  $\Z_{2mN_c,{{\bf Q} - N_c {\bf L}}}$.
Including 
the
Lorentz spacetime Spin group symmetry, we have
a  
faithful symmetry
for the gauge-invariant baryons:
\bea \label{eq:SpinB-LB+L}
\Spin \times_{\Z_2^\rF} \Z_{2m,{\bf B} - {\bf L}} \times_{\Z_2^\rF}\Z_{2 N_f, {\bf B} + {\bf L}}.
\eea
and for the free quarks:
\bea \label{eq:SpinQ-NcLQ+NcL}
\Spin \times_{\Z_2^\rF} \Z_{2m N_c,{\bf Q} - N_c {\bf L}} \times_{\Z_2^\rF}\Z_{2 N_c N_f, {\bf Q} + N_c {\bf L}}.
\eea
\end{enumerate}
Nonperturbative global anomalies of these discrete internal symmetries of (3+1)d fermions are topics of long-term interests, see for example
\cite{Ibanez1991,Preskill1991,Banks1992, 1808.00009,
1808.02881, 
GuoJW1812.11959,
WW2019fxh1910.14668}.
In particular, the cobordism approach
\cite{Kapustin1406.7329, 2016arXiv160406527F}
teaches us that
we should look at
$$
\left\{
\begin{array}{l}
\text{nonperturbative global anomalies in (3+1)d
classified by $\Z_p$
via the torsion part of bordism group
$(\Omega_5^G)_{\text{torsion}}$}.
\\
\text{perturbative local anomalies in (3+1)d
classified by $\Z$
via the free part of bordism group
$(\Omega_6^G)_{\text{free}}$}
.
\end{array}
\right.
$$
Let us explain the bordism group notation in depth in the next subsection \Sec{sec:Notations}.
Next we will summarize the anomaly index
of Weyl fermions within the bordism or TP group classification in
\Sec{sec:Anomaly-Index-bordism}.
We then summarize in terms of the Tables for the key anomaly indices of Weyl fermions and Symmetry-Extension Trivialization
in \Sec{sec:Tables}.

{The plan of this article goes as follows.}

In \Sec{sec:approach}, we outline the approach of this work, based on the computation of the Atiyah-Patodi-Singer $\eta$-invariants \cite{Atiyah1975jfAPS, Atiyah1976APS, Atiyah1980APS}.

In \Sec{sec:SManomalies}, we compute the anomaly index of the Standard Model (SM) or the anomaly index of (missing) sterile neutrinos involving the mixed gauge-gravitational anomaly of the discrete $({\bf B \pm L})$ internal symmetry and the gravitational background.

In \Sec{sec:SEexamples}, we study some examples of symmetry extension
 to trivialize fermionic anomalies by replacing the Weyl fermion theory with a (3+1)d $G$-symmetric anomalous 
 $K$-gauge topological order.
 
 In \Sec{sec:theoremproof}, we state and prove some general statements
on the symmetry-extension trivialization by $K$
for fermionic anomalies in $G$ of 
the 2-torsion,
the 3-torsion,
and the $s$-torsion nature. 

To recap the literature,
the symmetry-exntesion method is based on 
Witten \cite{Witten2016cio1605.02391}
and more systematically
Wang-Wen-Witten \cite{Wang2017locWWW1705.06728}.
There is some no go theorem about certain anomaly of a certain discrete symmetry cannot be matched by symmetric anomalous TQFTs, 
derived in Cordova-Ohmori \cite{Cordova2019bsd1910.04962, Cordova1912.13069}.
Microscopic condensed matter oriented construction of the (3+1)d anomalous symmetric fermionic TQFTs or topological orders are given in
\cite{Cheng:2024awi2411.05786}.
The categorical classification of (3+1)d symmetry enriched topological order with or without anomaly is recently proposed in D{\'e}coppet-Yu \cite{Decoppet:2025eic2509.10603}.
During the completion of this manuscript, in October, we becomes aware that
Debray-Ye-Yu \cite{Debray:2025kfg2510.24834} provides a different approach to solve the similar question:
``How to construct anomalous (3+1)d topological quantum field theories via symmetry extension approach?'' where Debray-Ye-Yu integrates the supercohomology and cobordism methods within the  categorical framework of fusion 2-categories \cite{Decoppet:2025eic2509.10603}. Instead, we will approach the similar question from the symmetry extension
trivialization of the Atiyah-Patodi-Singer eta invariant.
In addition, nonperturbative global anomalies of the standard model are explored recently in various works including
\cite{1808.00009, 1808.02881, DavighiGripaiosLohitsiri2019rcd1910.11277, WW2019fxh1910.14668, JW2006.16996, JW2008.06499, JW2012.15860,
WangWanYou2112.14765, WangWanYou2204.08393,
Wang:2025oow2502.21319}.

\subsection{Notations and Conventions of bordism group $\Omega$ 
and TP group}
\label{sec:Notations}

The group $\TP_{d+1}(G)$ is introduced by Freed-Hopkins in \cite{1604.06527}, classifying the anomalies of a $d$-dimensional quantum field theory with symmetry $G$. The elements of the group $\TP_{d+1}(G)$ are ($d+1$)-dimensional invertible topological quantum field theories with symmetry $G$. 
The group $\TP_{d+1}(G)$ is determined by the bordism groups $\Omega_{d+1}^G$ and $\Omega_{d+2}^G$ where
  \bea
  \Omega_d^G=\{\text{closed }d\text{-manifolds }M\text{ with }G\text{-structures on the tangent bundle }TM\}/\text{bordism}.
  \eea
It was shown in \cite{1604.06527} that there is a split short exact sequence 
\bea
0\to\text{Ext}^1(\Omega_d^G,\Z)\to\TP_d(G)\to\text{Hom}(\Omega_{d+1}^G,\Z)\to0.
\eea
Therefore, the torsion part of $\TP_d(G)$ is the same as the torsion part of $\Omega_d^G$, the free part of $\TP_d(G)$ is the same as the free part of $\Omega_{d+1}^G$. Namely,
\bea
\TP_d(G) = (\Omega_d^G)_{\text{torsion}} \oplus (\Omega_{d+1}^G)_{\text{free}}.
\eea
Since $\Omega_5^G$ and $\Omega_6^G$ are torsion for $G=\Spin\times\Z_n$ and $\Spin\times_{\Z_2^{\rF}}\Z_{2m}$, therefore $\TP_5(G)=\Omega_5^G$ for $G=\Spin\times\Z_n$ and $\Spin\times_{\Z_2^{\rF}}\Z_{2m}$.

\subsection{Summary of Anomaly Indices of
Weyl Fermions matching within the Bordism Group $\Omega$ or TP group} 

\label{sec:Anomaly-Index-bordism}

Because we mainly concern the (3+1)d fermionic theory with $G =\mathrm{Spin} \times_{\mathbb{Z}_2^{\rm F}} \mathbb{Z}_{2m}^{\rm F}$-symmetry,
we can factorize generically 
$$
\mathbb{Z}_{2m}^{\rm F}=
{\Z_{2^{p+1}  \cdot  3^r \cdot s
}}.
$$
We denote
${{\rm Spin}^{\Z_{2^{p+1}  \cdot  3^r \cdot s
}}} \equiv \frac{{\rm Spin} \times {\Z_{2^{p+1}  \cdot  3^r \cdot s
}}}{\Z_2^\rF}
\equiv {{\rm Spin} \times_{\Z_2^\rF} {\Z_{2^{p+1}  \cdot  3^r \cdot s
}}}$.
Here $p \geq 1$, $r \geq 1$, and $2 \nmid s$ and $3 \nmid s$.
The bordism group classifying the 4d  fermionic anomalies with symmetry ${\rm Spin}^{\Z_{2^{p+1}  \cdot  3^r \cdot s
}}$ is \cite{1808.02881}
\bea
&&\Omega_5^{{\rm Spin}^{\Z_{2^{p+1}  \cdot  3^r \cdot s
}}} 
\cong
\Omega_5^{{\rm Spin}^{\Z_{2^{p+1} 
}}}\oplus 
{\Omega}_5^{\rm Spin}(\B\Z_{3^r \cdot s}  )
\cong
\Omega_5^{{\rm Spin}^{\Z_{2^{p+1} 
}}}\oplus \tilde{\Omega}_5^{\rm SO}(\B\Z_{3^r \cdot s}  )
 \cr
&&=
\Z_{2^{p+3}}\oplus \Z_{2^{p-1}}\oplus
\Z_{3^{r+1}}\oplus \Z_{3^{r-1}}\oplus 
\Z_s \oplus 
\Z_s,
\text{ \quad $p \geq 1$, $r \geq 1$, and $2 \nmid s$ and $3 \nmid s$.}
\eea
Here $\tilde{\Omega}_5^{\rm SO}(\B G):=\Omega_5^{\rm SO}(\B G)/\Omega_5^{\rm SO}$ is the reduced bordism group, modding out the $\Omega_5^{\rm SO}=\Omega_5^{\rm SO}(pt)$.

We also concern the (3+1)d fermionic theory with $G =\mathrm{Spin} \times \mathbb{Z}_{n}$-symmetry,
we can factorize generically 
$$
\mathbb{Z}_{n}=
{\Z_{2^{p}  \cdot  3^r \cdot s
}}.
$$
Here $p \geq 2$, $r \geq 1$, and $2 \nmid s$ and $3 \nmid s$.
The bordism group classifying the 4d  fermionic anomalies with symmetry ${\rm Spin} \times {\Z_{2^{p}  \cdot  3^r \cdot s
}}$ is \cite{1808.02881}
\bea
&&\Omega_5^{{\rm Spin} \times {\Z_{2^{p}  \cdot  3^r \cdot s
}}} 
\cong
\Omega_5^{{\rm Spin}\times{\Z_{2^{p} 
}}}\oplus \Omega_5^{\Spin}(\B\Z_{3^r \cdot s}  )
\cong
\Omega_5^{{\rm Spin}\times{\Z_{2^{p} 
}}}\oplus \tilde{\Omega}_5^{\rm SO}(\B\Z_{3^r \cdot s}  )
 \cr
&&=
\Z_{2^{p}}\oplus \Z_{2^{p-2}}\oplus
\Z_{3^{r+1}}\oplus \Z_{3^{r-1}}\oplus 
\Z_s \oplus 
\Z_s,
\text{ \quad $p \geq 2$, $r \geq 1$, and $2 \nmid s$ and $3 \nmid s$.}
\eea

Here we focus on summarizing the anomaly \emph{indices}---not an explicit basis --- for
\bea
\text{$\TP_5(\Spin\times_{\Z_2^{\mathrm{F}}}\Z_{2m}^{\mathrm{F}})$
and
$\TP_5(\Spin\times\Z_n)$,}
\eea
those indices are obtained uniquely by reduction from the anomalies in
\bea
\text{
$\TP_5(\Spin\times\U(1))$ and 
$\TP_5(\Spin^c)$}.
\eea
The computation of these
indices are based on our approach in \Sec{sec:approach}.
Here we only list down the result as a summary for the convenience of checking anomaly indices. We also give two table summaries in
\Table{table-Spin-Zn}
and \Table{table-SpinxZn}.

The anomaly indices for $\TP_5(\Spin\times_{\Z_2^{\rF}}\Z_{2m})=\Z_{\tilde{a}_m}\oplus\Z_{\tilde{b}_m}$ are $\tilde{a}_m\frac{(2m^2+m+1)q^3-(m+3)q}{48m}$ and $\tilde{b}_m\frac{(m+1)\text{gcd}(m+1,2)(q^3-q)}{4m}$ \cite{2506.19710}.
The anomaly indices for $\TP_5(\Spin\times\Z_n)=\Z_{a_n}\oplus\Z_{b_n}$ are $\frac{a_n(n^2+3n+2)q^3}{6n}$ and $\frac{2b_n(q-q^3)}{n}$ \cite{2506.19710}.
In this article, we normalize the anomaly indices so that the anomaly indices for a single Weyl fermion with charge $q=1$ are $(1,0)$.

In particular, the anomaly index for $\TP_5(\Spin\times_{\Z_2^{\rF}}\Z_4)=\Z_{16}$ is $\frac{11q^3-5q}{6}\mod16$.

The anomaly index for $\TP_5(\Spin\times_{\Z_2^{\rF}}\Z_6)=\Z_{9}$ is $\frac{22q^3-6q}{16}\mod9$. 

The anomaly indices for $\TP_5(\Spin\times_{\Z_2^{\rF}}\Z_8)=\Z_{32}\oplus\Z_2$ are $\frac{37q^3-7q}{6}\mod32$ and $\frac{5(q^3-q)}{8}\mod2$. After normalization, the anomaly indices are $13\cdot\frac{37q^3-7q}{6}\mod32$ and $\frac{5(q^3-q)}{8}\mod2$.

The anomaly index for $\TP_5(\Spin\times_{\Z_2^{\rF}}\Z_{12})=\Z_{144}=\Z_{16}\oplus\Z_9$ is $\frac{79q^3-9q}{2}\mod144$. After normalization, the anomaly index is $-37\cdot\frac{79q^3-9q}{2}\mod144$.

The anomaly indices for $\TP_5(\Spin\times_{\Z_2^{\rF}}\Z_{18})=\Z_{27}
\oplus\Z_3$ are $\frac{172q^3-12q}{16}\mod27$ and $\frac{5(q^3-q)}{3}\mod3$. After normalization, the anomaly indices are $-8\cdot\frac{172q^3-12q}{16}\mod27$ and $\frac{5(q^3-q)}{3}\mod3$.

The anomaly index for $\TP_5(\Spin\times\Z_3)=\Z_9$ is $10q^3=q^3\mod9$.

The anomaly index for $\TP_5(\Spin\times\Z_4)=\Z_4$ is $5q^3=q^3\mod4$. 

The anomaly indices for $\TP_5(\Spin\times\Z_8)=\Z_8\oplus\Z_2$ are $15q^3=-q^3\mod8$ and $\frac{q-q^3}{2}\mod2$. After normalization, the anomaly indices are $q^3\mod8$ and $\frac{q-q^3}{2}\mod2$. 

The anomaly indices for $\TP_5(\Spin\times\Z_9)=\Z_{27}\oplus\Z_3$ are $55q^3=q^3\mod27$ and $\frac{2(q-q^3)}{3}\mod3$.

The anomaly indices for $\TP_5(\Spin\times\Z_{16})=\Z_{16}\oplus\Z_4$ are $51q^3=3q^3\mod16$ and $\frac{q-q^3}{2}\mod4$. After normalization, the anomaly indices are $q^3\mod16$ and $\frac{q-q^3}{2}\mod4$. 

\begin{table}[!h]
\begin{tabular}{| c  | c | c |  c |  c |  c | c | c |  c |  c |  c | }
\hline
Symmetry &  $q$ &  1  & 3 &  5  & 7 & 9 & 11 & 13 & 15 & 17\\
\hline
\hline
$\mathrm{Spin} \times_{\mathbb{Z}_2^{\rm F}} \mathbb{Z}_{4}^{\rm F}$ & $\tilde{\nu}_1'(m=2,q) \mod 16$ &$1$& $-1$ &  & &  &&  &&  \\
\hline
\hline
$\mathrm{Spin} \times_{\mathbb{Z}_2^{\rm F}} \mathbb{Z}_{6}^{\rm F}$ & $\tilde{\nu}_1'(m=3,q) \mod 9$ & $1$ & $0$ & $-1$ &  & &  & &  &\\
\hline
\hline
$\mathrm{Spin} \times_{\mathbb{Z}_2^{\rm F}} \mathbb{Z}_{8}^{\rm F}$ & $\tilde{\nu}_1'(m=4,q) \mod 32$ & $1$ & $7$ & $-7$ & $-1$ & &  & &  &\\
\hline
as above & $\tilde{\nu}_2'(m=4,q) \mod 2$ & $0$ & $1$ & $1$ & $0$ & &  & &  &\\
\hline
\hline
$\mathrm{Spin} \times_{\mathbb{Z}_2^{\rm F}} \mathbb{Z}_{12}^{\rm F}$ & 
$\tilde{\nu}_1'(m=6,q) \mod 144$ & $1$ & $63$ & $17$ & $-17$ & $-63$ &  $-1$ & &  &\\
\hline
as above & 
$\tilde{\nu}_{1,1}'(m=6,q) \mod 16$ & $1$ & $-1$ & $1$ & $-1$ & $1$ & $-1$  & &  &\\
\hline
as above & 
$\tilde{\nu}_{1,2}'(m=6,q) \mod 9$ & $1$ & $0$ & $-1$ & $1$ & $0$ & $-1$  & &  &\\
\hline
\hline
$\mathrm{Spin} \times_{\mathbb{Z}_2^{\rm F}} \mathbb{Z}_{18}^{\rm F}$ & 
$\tilde{\nu}_1'(m=9,q) \mod 27$ & $1$ & $-9$ & $-1$ & $1$ & $0$ & $-1$ & $1$ & $9$ & $-1$\\
\hline
as above & $\tilde{\nu}_2'(m=9,q) \mod 3$ & $0$ & $1$ & $2$ & $2$& $0$& $1$ & $1$ & $2$ & $0$\\
\hline
\end{tabular}
\caption{The anomaly index $\tilde{\nu}'$
of a charge-$q$ Weyl fermion $\psi_q$ within the symmetry
$\mathrm{Spin} \times_{\mathbb{Z}_2^{\rm F}} 
\mathbb{Z}_{2m}^{\rm F}$ where $q \in$ odd integers in $\mathbb{Z}_{2m}$.
Since $\TP_5(\Spin\times_{\Z_2^{\rF}}\Z_{2m})=\Omega_5^{\Spin\times_{\Z_2^{\rF}}\Z_{2m}}=\Z_{\tilde{a}_m}\oplus\Z_{\tilde{b}_m}$,
we have 
$(\tilde{\nu}_1'(m,q) \mod {\tilde{a}_m} )\in \Z_{\tilde{a}_m}$
and $(\tilde{\nu}_2'(m,q) \mod {\tilde{b}_m} ) \in \Z_{\tilde{b}_m}$. In the case of 
$\TP_5(\Spin\times_{\Z_2^{\rF}}\Z_{12})=\Omega_5^{\Spin\times_{\Z_2^{\rF}}\Z_{12}}=\Z_{144}
= \Z_{16}\oplus\Z_{9}$,
we break down
$(\tilde{\nu}_1'(6,q) \mod {144} )\in \Z_{144}$
as $(\tilde{\nu}_{1,1}'(6,q) \mod {16} )\in \Z_{16}$
and $(\tilde{\nu}_{1,1}'(9,q) \mod {9} )\in \Z_{9}$.
}
\label{table-Spin-Zn}
\end{table}

\begin{table}[!h]
\begin{tabular}{|c |c |  c   |  c |c   |  c |c |  c   |  c |c   |  c|c   |  c |c |  c   |  c |c   |  c | }
\hline
Symmetry & $q$ &0&1& 2 &3&4&5&  6&7&8&9 & 10 &11&12&13& 14&15 \\
\hline
\hline
$\mathrm{Spin} \times \mathbb{Z}_{3}$ &  $\nu_1'(n=3, q) \mod 9$ & 0 & 1 & $-1$ & 
&&&&&&&&&&&&\\
\hline
\hline
$\mathrm{Spin} \times \mathbb{Z}_{4}$ &  $\nu_1'(n=4, q) \mod 4$ & 0 & 1 & 0 & $-1$ 
&&&&&&&&&&&&\\
\hline
\hline
$\mathrm{Spin} \times \mathbb{Z}_{8}$ & 
$\nu_1'(n=8,q)\mod 8$ & 0 & $1$ & 0 & $3$ &  0 & $5$  & 0  & $7$ &&&&&&&&\\
\hline
as above & 
$\nu_2'(n=8,q)\mod 2$ & 0 & $0$ & $1$ & $0$ &  0 & $0$  & $1$  & $0$ &&&&&&&&\\
\hline
\hline
$\mathrm{Spin} \times \mathbb{Z}_{9}$ &  $\nu_1'(n=9, q) \mod 27$ & $0$ &  $1$ & $8$ & $0$ & $10$ & $-10$ & $0$ & $-8$ & $-1$ & & & & & & &
\\
\hline
as above & $\nu_2'(n=9, q) \mod 3$ & $0$ & $0$ & $2$ &  $2$ & $2$ & $1$ & $1$ & $1$ & $0$ & & & & & & &
\\
\hline
\hline
$\mathrm{Spin} \times \mathbb{Z}_{16}$ & $\nu_1'(n=16,q)\mod 16$ &0&1&8&11&0&13&8&7&0&9&8&3&0&5&8&15 \\
\hline
as above & $\nu_2'(n=16, q) \mod 4$ & 0& 0& 1& 0& 2& 0& 3& 0& 0&0&1&0&2&0&3&0 \\
\hline
\end{tabular}
\caption{The anomaly index ${\nu}'$
of a charge-$q$ Weyl fermion $\psi_q$ within the symmetry
$\mathrm{Spin} \times  
\Z_n$ where $q \in \mathbb{Z}_{n}$.
Since 
$\TP_5(\Spin\times\Z_n)=\Omega_5^{\Spin\times\Z_n}=\Z_{a_n}\oplus\Z_{b_n}$,
we have 
$({\nu}_1'(n,q) \mod {{a}_n} )\in \Z_{{a}_n}$
and $({\nu}_2'(n,q) \mod {{b}_n} ) \in \Z_{{b}_n}$. 
Although the charge-$q$ Weyl fermion in $\mathrm{Spin} \times \mathbb{Z}_{9}$ symmetry should have the same anomaly index as the charge-$q$ Weyl fermion  $\mathrm{Spin} \times_{\mathbb{Z}_2^{\rm F}} \mathbb{Z}_{18}^{\rm F}$ in the same basis, here because we choose a different basis, the anomaly indices are different for $\mathrm{Spin} \times \mathbb{Z}_{9}$ and 
$\mathrm{Spin} \times_{\mathbb{Z}_2^{\rm F}} \mathbb{Z}_{18}^{\rm F}$.
The anomaly indices would be mapped to each other via a linear map, namely 
$ \frac{1}{27}\tilde{\nu}_1'(m=9,q)+\frac{1}{3}\tilde{\nu}_2'(m=9,q)=\frac{1}{27}\nu_1'(n=9, q\mod 9) $
and 
$ \tilde{\nu}_2'(m=9,q)=-\nu_2'(n=9, q\mod 9) $
for odd $q\in\Z_{18}$.
}
\label{table-SpinxZn}
\end{table}

\newpage

\subsection{Table Summary of Anomaly Indices of Weyl Fermions and Symmetry-Extension Trivialization}

\label{sec:Tables}

\Table{table:Weyl-fermion-ext-1}
and 
\Table{table:Weyl-fermion-ext-2} show 
the symmetry extension trivialization 
of some anomalous
Weyl fermions with an anomaly index
$$
\nu_G \in \TP_d(G)
$$ 
becomes trivial by pulling back $G$ to
$G_{\rm Tot}$
so
$$
(r^*\nu)_{G_{\rm Tot}} =0 \in \TP_d(G_{\rm Tot})
$$
via the group extension short exact sequence:
\bea
1 \to K \to G_{\rm Tot} \to G \to 1.
\eea
Notice that for the 2-torsion kind of symmetry in $G$, we can use either
$K=\Z_4$ or 
$K=\Z_2$ to trivialize the $\nu_G \in \TP_d(G)$
anomaly.

\begin{table}[!h]
\begin{tabular}{| c |  c |  c  | c | c |  c |  c | }
\hline
$G$  & \text{ Anomaly class } & \text{ Anomaly index }&  $G_{\rm Tot}$  &  $K$ & \text{New anomaly index in $G_{\rm Tot}$} & Trivialized\\
\hline
{\parbox[c][3em][c]{4cm}{${\Spin \times_{\Z_2^\rF} {\Z_{2^{k+1}}^\rF}}$,
\\
for $k\geq 1$
}}
& 
\multirow{2}{*}{$\Z_{2^{k+3}} \oplus \Z_{2^{k-1}}$} & 
{\parbox[c][3em][c]{4cm}{
1 $\psi_{\rm W}$ of $q=1$
}} & 
{${\Spin \times {\Z_{2^{k+2}}^\rF}}$} 
& $\Z_4$ & 
{\parbox[c][5em][c]{4cm}{
$(\upnu_1,\upnu_2) = (\frac{4(2^{k+2}+1)(2^{k+2}+2)}{3},-3)$\\
$\in \Z_{2^{k+2}} \oplus \Z_{2^k}$.\\
1 $\psi_{\rm W}$ of $q'=2$
}} & {\parbox[c][3em][c]{2cm}{
no, but \\
yes for $2^k$ $\psi_{\rm W}$}}  \\
\cline{3-7}
& 
& 
{\parbox[c][3em][c]{4cm}{
1 $\psi_{\rm W}$ of $q=1$
}} & ${\Spin \times {\Z_{2^{k+1}}^\rF}}$ & $\Z_2$ & 
{\parbox[c][5em][c]{4cm}{
$(\upnu_1,\upnu_2) = (\frac{(2^{k+1}+1)(2^{k+1}+2)}{6},0)$\\
$\in \Z_{2^{k+1}} \oplus \Z_{2^{k-1}}$.\\
1 $\psi_{\rm W}$ of $q'=1$
}} & {\parbox[c][3em][c]{2cm}{
no, but \\
yes for $2^{k+1}$ $\psi_{\rm W}$}}  \\
\hline
{\parbox[c][3em][c]{4cm}{${\Spin \times 
{\Z_{2^{k+1}}^\rF}}$,
\\
for $k\geq 1$
}}
& 
{$\Z_{2^{k+1}} \oplus \Z_{2^{k-1}}$} & 
{\parbox[c][3em][c]{4cm}{
1 $\psi_{\rm W}$ of $q=1$
}} & ${\Spin \times {\Z_{2^{k+2}}^\rF}}$ & $\Z_2$ & 
{\parbox[c][5em][c]{4cm}{
$(\upnu_1,\upnu_2) =$  \\
$(\frac{4(2^{k+2}+1)(2^{k+2}+2)}{3},-3)$ \\
$\in \Z_{2^{k+2}} \oplus \Z_{2^k}$.\\
1 $\psi_{\rm W}$ of $q'=2$
}} & {\parbox[c][3em][c]{2cm}{
no, but \\
yes for $2^k$ $\psi_{\rm W}$}}  \\
\hline
\end{tabular}
\caption{The anomaly index of a Weyl fermion $\psi_{\rm W}$ of $q=1$ in a given $G$ symmetry,
and its minimal $K$ to succeed in the symmetry-extension trivialization
$1 \to K \to 
G_{\rm Tot} \to G \to 1$.
Here, a single $\psi_{\rm W}$ of $q=1$'s
anomaly index is not necessarily normalized as $(1,0)$
in either $\Omega^{\Spin \times_{\Z_2^\rF} {\Z_{2^{k+1}}^\rF}}_5=
{\Z_{2^{k+3}} \oplus \Z_{2^{k-1}}}$
or 
$\Omega^{\Spin \times_{} {\Z_{2^{k+1}}^\rF}}_5=
{\Z_{2^{k+1}} \oplus \Z_{2^{k-1}}}$; this is due to the choice basis.
We can replace the $G$-symmetric 4d Weyl fermion by
anomalous $G$-symmetric discrete-$K$-gauge 4d TQFTs
as the alternative 
low-energy theories matching the same symmetry and the same anomaly in the same deformation class.
}
\label{table:Weyl-fermion-ext-1}
\end{table}

\begin{table}[!h]
\begin{tabular}{| c |  c |  c  | c | c |  c |  c | }
\hline
$G$  & \text{ Anomaly class } & \text{ Anomaly index }&  $G_{\rm Tot}$  &  $K$ & \text{New anomaly index in $G_{\rm Tot}$} & Trivialized\\
\hline
\multirow{4}{*}{${\Spin \times_{\Z_2^\rF} {\Z_4^\rF}}$} & 
\multirow{4}{*}{$\Z_{16}$} & 
{\parbox[c][3em][c]{4cm}{
$\upnu = 1 \in \Z_{16}$.\\
1 $\psi_{\rm W}$ of $q=1$
}} & $\Spin \times {\Z_8^\rF}$ & $\Z_4$ & 
{\parbox[c][3em][c]{4cm}{
$\upnu = 1 \times (0,1) \in \Z_{8} \oplus \Z_{2}$.\\
1 $\psi_{\rm W}$ of $q'=2$
}} & no   \\
\cline{3-7}
& &  {\parbox[c][3em][c]{4cm}{
$\upnu = 2 \in \Z_{16}$.\\
2 $\psi_{\rm W}$ of $q=1$
}} & $\Spin \times {\Z_8^\rF}$ & $\Z_4$ & 
{\parbox[c][3em][c]{4cm}{
$\upnu = 2 \times (0,1) \in \Z_{8} \oplus \Z_{2}$.\\
2 $\psi_{\rm W}$ of $q'=2$
}} &  yes
\\
\cline{3-7}
& &  {\parbox[c][3em][c]{4cm}{
$\upnu = 4 \in \Z_{16}$.\\
4 $\psi_{\rm W}$ of $q=1$
}} & $\Spin \times {\Z_4^\rF}$  &  $\Z_2$  &  {\parbox[c][3em][c]{4cm}{
$\upnu = 4 \times 1 \in \Z_{4} $.\\
4 $\psi_{\rm W}$ of $q'=1$
}} &  yes \\
\cline{3-7}
& &  {\parbox[c][3em][c]{4cm}{
$\upnu = 8 \in \Z_{16}$.\\
8 $\psi_{\rm W}$ of $q=1$
}} & $\Spin \times {\Z_4^\rF}$ &  $\Z_2$ &  {\parbox[c][3em][c]{4cm}{
$\upnu = 8 \times 1 \in \Z_{4} $.\\
8 $\psi_{\rm W}$ of $q'=1$
}} &  yes\\
\hline
\multirow{5}{*}{${\Spin \times_{\Z_2^\rF} {\Z_8^\rF}}$} 
& 
\multirow{5}{*}{$\Z_{32}\oplus \Z_2$}
& 
{\parbox[c][3em][c]{4cm}{
{$(\upnu_1,\upnu_2) = (1,0) \in \Z_{32} \oplus \Z_2$}.\\
1 $\psi_{\rm W}$ of $q=1$
}}
& $\Spin \times {\Z_{16}^\rF}$ & $\Z_4$ &  {\parbox[c][3em][c]{5cm}{
$(\upnu_1,\upnu_2) = 1 \times (8,1) \in \Z_{16} \oplus \Z_{4}$.\\
1 $\psi_{\rm W}$ of $q'=2$
}} & 
{\parbox[c][3em][c]{2cm}{
no, but \\
yes for 4 $\psi_{\rm W}$}}
\\
\cline{3-7}
&&{\parbox[c][3em][c]{4cm}{
{$(\upnu_1,\upnu_2) = (7,1) \in \Z_{32} \oplus \Z_2$}.\\
1 $\psi_{\rm W}$ of $q=3$
}}
&$\Spin \times {\Z_{16}^\rF}$ & $\Z_4$ &  {\parbox[c][3em][c]{5cm}{
$(\upnu_1,\upnu_2) = 1 \times (8,3) \in \Z_{16} \oplus \Z_{4}$.\\
1 $\psi_{\rm W}$ of $q'=6$
}} & 
{\parbox[c][3em][c]{2cm}{
no, but \\
yes for 4 $\psi_{\rm W}$}}\\
\cline{3-7}
&&{\parbox[c][3em][c]{5cm}{
{$(\upnu_1,\upnu_2) = -(7,1) \in \Z_{32} \oplus \Z_2$}.\\
1 $\psi_{\rm W}$ of $q=5$
}}
&$\Spin \times {\Z_{16}^\rF}$ & $\Z_4$ &  {\parbox[c][3em][c]{5cm}{
$(\upnu_1,\upnu_2) = 1 \times (8,1) \in \Z_{16} \oplus \Z_{4}$.\\
1 $\psi_{\rm W}$ of $q'=10$
}} & 
{\parbox[c][3em][c]{2cm}{
no, but \\
yes for 4 $\psi_{\rm W}$}}
\\
\cline{3-7}
&&{\parbox[c][3em][c]{5cm}{
{$(\upnu_1,\upnu_2) = -(1,0) \in \Z_{32} \oplus \Z_2$}.\\
1 $\psi_{\rm W}$ of $q=7$
}}
&$\Spin \times {\Z_{16}^\rF}$ & $\Z_4$ &  {\parbox[c][3em][c]{5cm}{
$(\upnu_1,\upnu_2) = 1 \times (8,3) \in \Z_{16} \oplus \Z_{4}$.\\
1 $\psi_{\rm W}$ of $q'=14$
}} & 
{\parbox[c][3em][c]{2cm}{
no, but \\
yes for 4 $\psi_{\rm W}$}}
\\
\cline{3-7}
&&{\parbox[c][4em][c]{5cm}{
{$(\upnu_1,\upnu_2) = (0,1) \in \Z_{32} \oplus \Z_2$}.\\
7 $\psi_{\rm W}$ of $q=-1$, \\
1 $\psi_{\rm W}$ of $q=3$.
}}
&$\Spin \times {\Z_{16}^\rF}$ & $\Z_4$ &  {\parbox[c][4em][c]{5cm}{
$(\upnu_1,\upnu_2) = 8 \times (8,3) \in \Z_{16} \oplus \Z_{4}$.\\
7 $\psi_{\rm W}$ of $q'=-2$, \\
1 $\psi_{\rm W}$ of $q'=6$.
}} & yes \\
\hline
\end{tabular}
\caption{The anomaly index of some Weyl fermions $\psi_{\rm W}$ of various $q$ in a given $G$ symmetry,
and its minimal $K$ to succeed in the symmetry-extension trivialization
$1 \to K \to 
G_{\rm Tot} \to G \to 1$.
We can replace the $G$-symmetric 4d Weyl fermion by
anomalous $G$-symmetric discrete-$K$-gauge 4d TQFTs
as the alternative 
low-energy theories matching the same symmetry and the same anomaly in the same deformation class.
}
\label{table:Weyl-fermion-ext-2}
\end{table}

\newpage
\section{Approach: From
Atiyah-Patodi-Singer $\eta$-invariants to Symmetry-Extension Trivialization
}\label{sec:approach}

In this section, we outline the approach of this work, based on the computation of the Atiyah-Patodi-Singer $\eta$-invariants \cite{Atiyah1975jfAPS, Atiyah1976APS, Atiyah1980APS, Witten2019bou1909.08775},
and their symmetry-extension trivialization of the anomaly indices.

\begin{enumerate}

\item The anomalies
    of 4d Weyl fermions with symmetries $G=\Spin\times\Z_n$ or $G=\Spin\times_{\Z_2^{\rF}}\Z_{2m}$ that 
    contains internal discrete finite symmetries 
    with their representations $\rho$
    are expressed
    in terms of an eta invariant
    \bea
    \eta(M^5,\rho) \in \R/\Z ,
    \eea
    and its 5d invertible field theory on a 5-manifold $M^5$,
    \bea
    \exp(-2\pi\ii\eta(M^5,\rho)) \in
    \exp(-2\pi\ii\R/\Z) \equiv \U(1) 
    \eea
where $\rho$ is a representation of $\Z_n$ or $\Z_{2m}$ in our cases. Following \cite{DaiFreed1994}, the partition function of a closed 4-manifold $M^4$ and a compact 5-manifold $M^5$ with boundary $\partial M^5=M^4$ is
\bea
Z_{\psi}=|\det(\cD(M^4,\rho))|\exp(-2\pi\ii \eta(M^5,\rho))
\eea
where $\cD(M^4,\rho)$ is the Dirac operator. Here, $\det(\cD(M^4,\rho))$ is the Gaussian fermion path integral of 
\bea
\int [D \bar{\psi}] [D\psi]
\exp(-\bar{\psi}\cD(M^4,\rho)\psi) =
\det(\cD(M^4,\rho))
\eea
integrating out the Grassmann number $\psi$. 
Following \cite{Witten1508.04715,Witten2016cio1605.02391,1607.01873,Witten2019bou1909.08775}, $\exp(-2\pi\ii \eta(M^5,\rho))$ is defined as the partition function over the domain wall
with a positive mass 
$|m_0| \to +\infty$ 
and
a negative mass
$-|m_0| \to -\infty$ along the 5th dimension: 
\bea
\exp(-2\pi\ii \eta(M^5,\rho)):=\lim_{m_0\to\infty}\frac{\det(-\ii \cD(M^5,\rho)-m_0)}{\det(-\ii \cD(M^5,\rho)+m_0)}=\lim_{m_0\to\infty}\prod_{\lambda_{M^5}}\frac{-\ii\lambda_{M^5}-m_0}{-\ii\lambda_{M^5}+m_0}
\eea
where $\lambda_{M^5}$ are the eigenvalues of $\cD(M^5,\rho)$.
We define $s(\lambda_{M^5})\in(-1,1]$ by 
\bea
\exp(-\pi\ii s(\lambda_{M^5})):=\frac{-\ii\lambda_{M^5}-m_0}{-\ii\lambda_{M^5}+m_0}=\frac{-m_0^2+\lambda_{M^5}^2-2m_0\lambda_{M^5}\ii}{m_0^2+\lambda_{M^5}^2}.
\eea
Since $\exp(-\pi\ii s(\lambda_{M^5}))=\cos(\pi s(\lambda_{M^5}))-\ii\sin(\pi s(\lambda_{M^5}))$, we have $s(0)=1$. \\
If $\lambda_{M^5}>0$, then $s(\lambda_{M^5})\to 1$ from below $s(\lambda_{M^5})\leq 1$, as $m_0\to\infty$. \\
If $\lambda_{M^5}<0$, then $s(\lambda_{M^5})\to -1$ from above $s(\lambda_{M^5}) > -1$, as $m_0\to\infty$.\\ 
Therefore, if $\lambda_{M^5}\ne0$, then $s(\lambda_{M^5})\to \text{sign}(\lambda_{M^5})$ as $m_0\to\infty$.\\
If $\frac{\lambda_{M^5}}{m_0} \to\pm\infty$, $s(\lambda_{M^5})\to0$.

By the definitions, 
\bea
\exp(-2\pi\ii \eta(M^5,\rho))=\lim_{m_0\to\infty}\exp(-\pi\ii\sum_{\lambda_{M^5}}s(\lambda_{M^5})).
\eea
Therefore, 
\bea
\eta(M^5,\rho)=\frac{1}{2}(\sum_{\lambda_{M^5}\ne0}\text{sign}(\lambda_{M^5})+\dim\ker \cD(M^5,\rho)),
\eea
where $\dim\ker \cD(M^5,\rho)$ counts the number of zero eigenvalues of the Dirac operator $\cD(M^5,\rho)$.
However, the sum $\sum_{\lambda_{M^5}\ne0}\text{sign}(\lambda_{M^5})$ is not well-defined mathematically. 
The regularized function of $s$
\bea
\eta(s):=\sum_{\lambda_{M^5}\ne0}\text{sign}(\lambda_{M^5})|\lambda_{M^5}|^{-s}
\eea
converges for $\text{Re }s$ large and admits analytic continuation to $s=0$.
The regularized $\eta$-invariant is 
\bea\label{eq:eta-invariant}
\eta(M^5,\rho)=\frac{1}{2}(\eta(0)+\dim\ker \cD(M^5,\rho)),
\eea
which takes value in $\R/\Z$.

    \item
In \cite{1808.02881}, Hsieh determined the group structure of $\TP_5(G)$ for $G=\Spin\times\Z_n$ or $G=\Spin\times_{\Z_2^{\rF}}\Z_{2m}$. Namely,
\bea\label{eq:Spin-Zn-TP5}
\TP_5(\Spin\times\Z_n)&=&\Omega_5^{\Spin\times\Z_n}=\Z_{a_n}\oplus\Z_{b_n} \equiv \left\{\begin{array}{llll}0&n=2,\\\Z_n\oplus\Z_{\frac{n}{4}}&n=2^{\nu},\;\nu>1,\\\Z_{3n}\oplus\Z_{\frac{n}{3}}&n=3^{\nu},\\\Z_n\oplus\Z_n&n=p^{\nu},\;p>3\text{ is prime},\end{array}\right.
\eea
    and 
    \bea\label{eq:Spin-Z2m-TP5}
\TP_5(\Spin\times_{\Z_2^{\rF}}\Z_{2m})&=&\Omega_5^{\Spin\times_{\Z_2^{\rF}}\Z_{2m}}=\Z_{\tilde{a}_m}\oplus\Z_{\tilde{b}_m}\equiv \left\{\begin{array}{llll}0&m=1,\\\Z_{8m}\oplus\Z_{\frac{m}{2}}&m=2^{\nu}>1,\\\Z_{3m}\oplus\Z_{\frac{m}{3}}&m=3^{\nu},\\\Z_m\oplus\Z_m&m=p^{\nu},\;p>3\text{ is prime}.\end{array}\right.
\eea

    \item 

We will determine the anomaly indices for $\TP_5(\Spin\times\Z_n)$ and $\TP_5(\Spin\times_{\Z_2^{\rF}}\Z_{2m})$ in the following steps:
\begin{itemize}
    \item First, for $G=\Spin\times\Z_n$ and $G=\Spin\times_{\Z_2^{\rF}}\Z_{2m}$, we find two 5-manifolds with  $G$-structures respectively.

    For $G=\Spin\times\Z_n$, we find two 5-manifolds with $G$-structures: $X(n;1,1)$ and $L(n;1)\times\text{K3}$.
Here,
\begin{itemize}
    \item 
\bea
X(n;1,1)=S(H\otimes H\oplus 1)/\tau(1,1)
\eea
where $H$ is the Hopf line bundle over $S^2$, $1$ is the trivial complex line bundle over $S^2$, $S(H\otimes H\oplus 1)$ is the sphere bundle of the vector bundle $H\otimes H\oplus 1\cong H\oplus H$, and $\tau(1,1)$ is the diagonal action $\tau(1,1)(\lambda)=\diag(\lambda,\lambda)$ of $\Z_n=\{\lambda\in\C|\lambda^n=1\}$ on the sphere bundle $S(H\otimes H\oplus 1)$. $X(n;1,1)$ is a lens space bundle over $S^2$ with fiber the 3d lens space $L(n;1,1)=S^3/\tau(1,1)$. $X(n;1,1)$ has a natural $\Spin\times\Z_n$ structure. 
\item

$L(n;1)=S^1/\rho_1$, $\rho_1$ is the 1-dimensional representation $\rho_1(\lambda)=\lambda$ of $\Z_n=\{\lambda\in\C|\lambda^n=1\}$, and K3 is the 4d K3 (complex) surface. The signature of K3 is $-16$, hence the first Pontryagin class of K3 is $-48$. $L(n;1)\times\text{K3}$ has a natural $\Spin\times\Z_n$ structure. 
  \end{itemize}
    For $G=\Spin\times_{\Z_2^{\rF}}\Z_{2m}$ for even $m$, we find two 5-manifolds with $G$-structures: $L(m;1,1,1)$ and $L(m;1)\times\text{E}$.

    For $G=\Spin\times_{\Z_2^{\rF}}\Z_{2m}$ for odd $m$, we find two 5-manifolds with $G$-structures: $L(m;1,1,1)$ and $L(m;1)\times\text{K3}$. Since $\Spin\times_{\Z_2^{\rF}}\Z_{2m}=\Spin\times\Z_m$ for odd $m$, $L(m;1)\times\text{K3}$ has a $\Spin\times_{\Z_2^{\rF}}\Z_{2m}$ structure for odd $m$.

A $\Spin\times_{\Z_2^{\rF}}\Z_{2m}$ structure on a manifold $M$ is equivalent to a spin structure on $TM\oplus f^*\xi$ where $TM$ is the tangent bundle, $f:M\to \B\Z_m$ is a map and $\xi$ is the underlying real plane bundle of the complex line bundle determined by the natural map $\B\Z_m\to\B\U(1)$. An orientable manifold $M$ admits a $\Spin\times_{\Z_2^{\rF}}\Z_{2m}$ structure if and only if $w_2(TM)=w_2(f^*\xi)$ for some map $f$ where $w_2$ is the second Stiefel-Whitney class.

Here,
\begin{itemize}
    \item 
    $L(m;1,1,1)=S^5/\tau(1,1,1)$ is the 5d lens space where $\tau(1,1,1)$ is the diagonal action $\tau(1,1,1)(\lambda)=\diag(\lambda,\lambda,\lambda)$ of $\Z_m=\{\lambda\in\C|\lambda^m=1\}$ on $S^5\subset\C^3$. $L(m;1,1,1)$ has a natural $\Spin\times_{\Z_2^{\rF}}\Z_{2m}$ structure. 
    \item 
 $L(m;1)=S^1/\rho_1$, $\rho_1$ is the 1-dimensional representation $\rho_1(\lambda)=\lambda$ of $\Z_m=\{\lambda\in\C|\lambda^m=1\}$, and
    E is the 4d Enriques (complex) surface. The signature of E is $-8$, hence the first Pontryagin class of E is $-24$. In \cite{2506.19710}, Wan showed that E has a $\Spin\times_{\Z_2^{\rF}}\Z_{2m}$ structure for even $m$.
    
\end{itemize}

    \item Then, we compute the $\eta$-invariants of these manifolds.

    The anomalies in $\TP_5(\Spin\times\Z_n)$ are given by $\exp(-2\pi\ii\eta(M^5,\rho_q))$ where $\rho_q\in RU(\Z_n)$. Here, 
\bea
RU(\Z_n)=\bigoplus_{q=0}^{n-1}\rho_q\Z 
\eea
denotes the unitary representation ring of the group $\Z_n$, and $\rho_q(\lambda)=\lambda^q$ is a 1-dimensional representation of $\Z_n=\{\lambda\in\C|\lambda^n=1\}$. In terms of physics, for the representation $R$ of multiple fermions with charge $q$, the number of fermions with charge $q$ is the integer coefficient $\Z$ of $\rho_q$ in $R$.

The anomalies in $\TP_5(\Spin\times_{\Z_2^{\rF}}\Z_{2m})$ are given by $\exp(-2\pi\ii\eta(M^5,\tilde{\rho}_q))$ where $\tilde{\rho}_q\in RU^o(\Z_{2m})$. Here, $RU^o(\Z_{2m})$ denotes the additive subgroup of $RU(\Z_{2m})$ generated by the 1-dimensional representations $\tilde{\rho}_q$ of the group $\Z_{2m}$ with odd charges $q$.

The $\eta$-invariant $\eta(X(n;1,1),\rho_q)$ can be computed via \cite[Theorem 4.7]{Gilkey1997} and the Lefschetz fixed point formula. 
The $\eta$-invariant $\eta(L(m;1,1,1),\tilde{\rho}_q)$ can be computed via \cite[Theorem 4.6]{Gilkey1997} and the Lefschetz fixed point formula. See \cite{1808.02881} for the computation of these two $\eta$-invariants.

The $\eta$-invariants $\eta(L(n;1)\times\text{K3},\rho_q)$, $\eta(L(m;1)\times\text{E},\tilde{\rho}_q)$, and $\eta(L(m;1)\times\text{K3},\tilde{\rho}_q)$ can be computed via  \cite{1808.00009,Gilkey1996} and the fact \cite{Gilkey1987,Gilkey2018} that
\bea\label{eq:eta-product}
\eta(A\times B)=\eta(A)\hat{A}(B)
\eea
where $A$ is a 1-manifold, $B$ is a 4-manifold and $\hat{A}(B)$ is the $\hat{A}$-genus of $B$, see \cite{2506.19710} for the computation of these two $\eta$-invariants.

By \cite[Theorem 4.7]{Gilkey1997} and the Lefschetz fixed point formula, we have
\bea
\eta(X(n;1,1),\rho_q)=\frac{1}{n}\sum_{\lambda^n=1,\lambda\ne1}\lambda^q\frac{\lambda(1+\lambda)}{(1-\lambda)^3}.
\eea
On the other hand, by \cite[Theorem 4.6]{Gilkey1997} and the Lefschetz fixed point formula, we have 
\bea
\eta(L(n;1,1,1,1),R)=\frac{1}{n}\sum_{\lambda^n=1,\lambda\ne1}\text{Tr}(R(\lambda))\frac{\lambda^2}{(1-\lambda)^4}
\eea
where $L(n;1,1,1,1)$ is the 7d lens space $S^7/\tau(1,1,1,1)$ and $\tau(1,1,1,1)$ is the diagonal action $\tau(1,1,1,1)(\lambda)=\diag(\lambda,\lambda,\lambda,\lambda)$ of $\Z_n=\{\lambda\in\C|\lambda^n=1\}$ on $S^7$.
Comparing the above two formulas, we find that
\bea\label{eq:eta-X(n;1,1)}
\eta(-X(n;1,1),\rho_q)=\eta(L(n;1,1,1,1),(\rho_1-\rho_{-1})\rho_q).
\eea
Here and thereafter, the minus sign before a manifold indicates the reversal of its orientation.
By \cite{Gilkey1984}, we have
\bea
\eta(L(n;1,1,1,1),\rho_q)=-\frac{1}{n}\hat{A}_4(q+\frac{n}{2};n,1,1,1,1)\mod\Z,
\eea
where 
\bea
\hat{A}_k(t,\vec{x}):=\sum_{a+2b=k}\frac{t^a\hat{A}_b(\vec{x})}{a!}
\eea
and $\hat{A}_b(\vec{x})$ are the A-roof polynomials.
In \cite{1808.02881}, Hsieh computed $\eta(X(n;1,1),\rho_q)$ using the above method, the result is
\bea
\eta(X(n;1,1),\rho_q)=\frac{(n^2+3n+2)q^3}{6n}\mod\Z.
\eea

Since $\hat{A}(\text{K3})=-\frac{p_1}{24}=2$ where $p_1$ is the first Pontryagin class of K3, by \eqref{eq:eta-product} and \cite{1808.00009,Gilkey1996}, we have
\bea
\eta(L(n;1)\times\text{K3},\rho_q)=2\cdot\frac{1}{n}\sum_{\lambda^n=1,\lambda\ne1}(\lambda^q-1)\frac{\sqrt{\lambda}}{\lambda-1}.
\eea
In \cite{2506.19710}, Wan computed this $\eta$-invariant, the result 
on the orientation reversal of $L(n;1)\times\text{K3}$
is
\bea
\eta(-L(n;1)\times\text{K3},\rho_q)=\frac{2q}{n}\mod\Z.
\eea

By \cite[Theorem 4.6]{Gilkey1997} and the Lefschetz fixed point formula, we have
\bea
\eta(L(m;1,1,1),R)=\frac{1}{m}\sum_{\lambda^{m}=1,\lambda\ne1}\text{Tr}(R(\lambda))\frac{\lambda^2}{(1-\lambda)^3}
\eea
where $R=\bigoplus_qk_q\rho_{\frac{q-1}{2}}\in RU(\Z_m)$ corresponds to $\tilde{R}=\bigoplus_qk_q\tilde{\rho}_{q}\in RU^o(\Z_{2m})$.

On the other hand, by \cite[Theorem 4.6]{Gilkey1997} and the Lefschetz fixed point formula, we have 
\bea
\eta_{\Spin\times\Z_m}(L(m;1,1,1,1),R)=\frac{1}{m}\sum_{\lambda^m=1,\lambda\ne1}\text{Tr}(R(\lambda))\frac{\lambda^2}{(1-\lambda)^4}
\eea
where $L(m;1,1,1,1)$ is the 7d lens space $S^7/\tau(1,1,1,1)$  and $\tau(1,1,1,1)$ is the diagonal action $\tau(1,1,1,1)(\lambda)=\diag(\lambda,\lambda,\lambda,\lambda)$ of $\Z_m=\{\lambda\in\C|\lambda^m=1\}$ on $S^7$.
Comparing the above two formulas, we find that
\bea\label{eq:eta-L(m;1,1,1)}
\eta(L(m;1,1,1),\tilde{\rho}_q)=\eta_{\Spin\times\Z_m}(L(m;1,1,1,1),(\rho_0-\rho_{1})\rho_{\frac{q-1}{2}}).
\eea
Similarly, $\eta_{\Spin\times\Z_m}(L(m;1,1,1,1),(\rho_0-\rho_{1})\rho_{\frac{q-1}{2}})$ can be computed in terms of the A-roof polynomials. In \cite{1808.02881}, Hsieh computed $\eta(L(m;1,1,1),\tilde{\rho}_q)$ using the above method, the result is
\bea
\eta(L(m;1,1,1),\tilde{\rho}_q)=\frac{(2m^2+m+1)q^3-(m+3)q}{48m}\mod\Z.
\eea

Since $\hat{A}(\text{E})=-\frac{p_1}{24}=1$ where $p_1$ is the first Pontryagin class of E, by \eqref{eq:eta-product} and \cite{1808.00009,Gilkey1996,Gilkey2018}, we have
\bea
\eta(L(m;1)\times\text{E},\tilde{\rho}_q)=\frac{1}{2m}\sum_{\lambda^{2m}=1,\lambda\ne1}(\lambda^q-1)\frac{\lambda}{\lambda-1}.
\eea
In \cite{2506.19710}, Wan computed this $\eta$-invariant, the result on the orientation reversal of $L(m;1)\times\text{E}$ is
\bea
\eta(-L(m;1)\times\text{E},\tilde{\rho}_q)=\frac{q}{2m}\mod\Z.
\eea

Similarly, since $\hat{A}(\text{K3})=-\frac{p_1}{24}=2$ where $p_1$ is the first Pontryagin class of K3, by \eqref{eq:eta-product} and \cite{1808.00009,Gilkey1996,Gilkey2018}, we have
\bea
\eta(-L(m;1)\times\text{K3},\tilde{\rho}_q)=\frac{q}{m}\mod\Z.
\eea

For $G=\Spin\times\Z_n$, we denote the $\eta$-invariants as
    \bea
    \nu_1(n,q)&:=&\eta(X(n;1,1),\rho_q)=\frac{(n^2+3n+2)q^3}{6n}\mod\Z,\cr
    \nu_2(n,q)&:=&\eta(-L(n;1)\times\text{K3},\rho_q)=\frac{2q}{n}\mod\Z.
    \eea

    For $G=\Spin\times_{\Z_2^{\rF}}\Z_{2m}$, we denote the $\eta$-invariants as
    \bea
    \tilde{\nu}_1(m,q)&:=&\eta(L(m;1,1,1),\tilde{\rho}_q)=\frac{(2m^2+m+1)q^3-(m+3)q}{48m}\mod\Z,\cr
    \tilde{\nu}_2(m,q)&:=&\left\{\begin{array}{ll}\eta(-L(m;1)\times\text{E},\tilde{\rho}_q)=\frac{q}{2m}\mod\Z,&\text{for even }m\\
    \eta(-L(m;1)\times\text{K3},\tilde{\rho}_q)=\frac{q}{m}\mod\Z,&\text{for odd }m
    \end{array}\right.=\frac{\text{gcd}(m+1,2)q}{2m}.
    \eea

    \item We determine the generators of $\TP_5(\Spin\times\Z_n)$ and $\TP_5(\Spin\times_{\Z_2^{\rF}}\Z_{2m})$ by taking the integral linear combinations of these $\eta$-invariants.

$(\exp(-2\pi\ii\nu_1),\exp(-2\pi\ii\nu_2))$ are elements in $\TP_5(\Spin\times\Z_n)$, and $(\exp(-2\pi\ii\tilde{\nu}_1),\exp(-2\pi\ii\tilde{\nu}_2))$ are elements in $\TP_5(\Spin\times_{\Z_2^{\rF}}\Z_{2m})$.

However, $(a_n\nu_1,b_n\nu_2)$ are not necessarily the generators of $\TP_5(\Spin\times \Z_n)=\Z_{a_n}\oplus\Z_{b_n}$, and $(\tilde{a}_m\tilde{\nu}_1,\tilde{b}_m\tilde{\nu}_2)$ are not necessarily the generators of $\TP_5(\Spin\times_{\Z_2^{\rF}}\Z_{2m})=\Z_{\tilde{a}_m}\oplus\Z_{\tilde{b}_m}$. More precisely, $b_n\nu_2$ and $\tilde{b}_m\tilde{\nu}_2$ are not necessarily integer-valued. 

In Ref.~\cite{2506.19710}, for a charge $q$ left-handed Weyl fermion $\psi$, Wan determined the anomaly indices
\bea \label{eq:nu'}
(\nu_1'(n,q),\nu_2'(n,q)) = (\frac{a_n(n^2+3n+2)q^3}{6n} \mod a_n, \frac{2b_n(q-q^3)}{n}\mod b_n) \cr\text{ in }
\TP_5(\Spin\times\Z_n)=\Z_{a_n}\times\Z_{b_n}
\eea
and
\bea\label{eq:tildenu'}
(\tilde{\nu}_1'(m,q),\tilde{\nu}_2'(m,q))=(\tilde{a}_m\frac{(2m^2+m+1)q^3-(m+3)q}{48m} \mod \tilde{a}_m, \tilde{b}_m\frac{(m+1)\text{gcd}(m+1,2)(q^3-q)}{4m}\mod \tilde{b}_m)\cr\text{ in }\TP_5(\Spin\times_{\Z_2^{\rF}}\Z_{2m})=\Z_{\tilde{a}_m}\times\Z_{\tilde{b}_m}
\eea
such that $(\nu_1',\nu_2')$ are the generators of $\TP_5(\Spin\times \Z_n)=\Z_{a_n}\times\Z_{b_n}$, and $(\tilde{\nu}_1',\tilde{\nu}_2')$ are the generators of $\TP_5(\Spin\times_{\Z_2^{\rF}}\Z_{2m})=\Z_{\tilde{a}_m}\times\Z_{\tilde{b}_m}$.

Here,
 \bea
 \nu_1'(n,q)&=&a_n\eta(X(n;1,1),\rho_q),\cr
 \nu_2'(n,q)&=&b_n\eta(-L(n;1)\times\text{K3}-6X(n;1,1),\rho_q)
\eea
and
\bea
\tilde{\nu}_1'(m,q)&=&\tilde{a}_m\eta(L(m;1,1,1),\tilde{\rho}_q),\cr
\tilde{\nu}_2'(m,q)&=&\left\{\begin{array}{ll}\tilde{b}_m\eta(-L(m;1)\times\text{E}+12L(m;1,1,1),\tilde{\rho}_q)),&\text{for even }m\\
\tilde{b}_m\eta(-L(m;1)\times\text{K3}+24L(m;1,1,1),\tilde{\rho}_q)),&\text{for odd }m
\end{array}\right.
\eea
where the addition of manifolds means the disjoint union of manifolds.

By \eqref{eq:Spin-Zn-TP5}, \eqref{eq:nu'}, \eqref{eq:Spin-Z2m-TP5}, and \eqref{eq:tildenu'}, $(\nu_1',\nu_2')$ and $(\tilde{\nu}_1',\tilde{\nu}_2')$ are all integer-valued.

By \eqref{eq:Spin-Zn-TP5} and \eqref{eq:nu'}, for $q=1$, $\nu_1'(n,1)=\frac{a_n(n^2+3n+2)}{6n}$ is the generator of $\Z_{a_n}$ while $\nu_2'(n,1)=0\in\Z_{b_n}$, and for $q=2$, $\nu_2'(n,2)=\frac{-12b_n}{n}$ is the generator of $\Z_{b_n}$.
  So $(\nu_1',\nu_2')$ are the generators of $\TP_5(\Spin\times \Z_n)=\Z_{a_n}\times\Z_{b_n}$.

By \eqref{eq:Spin-Z2m-TP5} and \eqref{eq:tildenu'}, for $q=1$, $\tilde{\nu}_1'(m,1)=\frac{\tilde{a}_m(m^2-1)}{24m}$ is the generator of $\Z_{\tilde{a}_m}$ while $\tilde{\nu}_2'(m,1)=0\in\Z_{\tilde{b}_m}$, and for $q=3$, $\tilde{\nu}_2'(m,3)=\frac{6\tilde{b}_m(m+1)\text{gcd}(m+1,2)}{m}$ is the generator of $\Z_{\tilde{b}_m}$.
 So $(\tilde{\nu}_1',\tilde{\nu}_2')$ are the generators of $\TP_5(\Spin\times_{\Z_2^{\rF}}\Z_{2m})=\Z_{\tilde{a}_m}\times\Z_{\tilde{b}_m}$.

 Moreover, $(\nu_1',\nu_2')$ and $(\tilde{\nu}_1',\tilde{\nu}_2')$ are chosen so that $(\nu_1',\nu_2')$ are the images of the anomalies $qI_5^A+\frac{q^3-q}{6}I_5^B$ in $\TP_5(\Spin\times\U(1))=\Z^2$ under the natural group homomorphism from $\TP_5(\Spin\times\U(1))=\Z^2$ to $\TP_5(\Spin\times \Z_n)=\Z_{a_n}\times\Z_{b_n}$ where $I_5^A$ and $I_5^B$ are the generators of $\TP_5(\Spin\times\U(1))=\Z^2$, and $(\tilde{\nu}_1',\tilde{\nu}_2')$ are the images of the anomalies $q\tilde{I}_5^C+\frac{q^3-q}{24}\tilde{I}_5^D$ in $\TP_5(\Spin^c)=\Z^2$ under the natural group homomorphism from $\TP_5(\Spin^c)=\Z^2$ to $\TP_5(\Spin\times_{\Z_2^{\rF}}\Z_{2m})=\Z_{\tilde{a}_m}\times\Z_{\tilde{b}_m}$ where $\tilde{I}_5^C$ and $\tilde{I}_5^D$ are the generators of $\TP_5(\Spin^c)=\Z^2$, see \cite{2506.19710} for more explanation.
 
    In this article, we will use the anomaly indices $(\nu_1',\nu_2')$ for $\TP_5(\Spin\times \Z_n)$ and $(\tilde{\nu}_1',\tilde{\nu}_2')$ for $\TP_5(\Spin\times_{\Z_2^{\rF}}\Z_{2m})$.

\end{itemize}

\item While we have only computed the anomaly \emph{indices}---not an explicit basis---for
\(\TP_5(\Spin\times\Z_n)\) and \(\TP_5(\Spin\times_{\Z_2^{\mathrm{F}}}\Z_{2m})\),
those indices are obtained uniquely by reduction from the anomalies in
\(\TP_5(\Spin\times\U(1))\) (respectively \(\TP_5(\Spin^c)\)).
Under a symmetry extension
$
1 \to K \to G_{\rm Tot} \to G \to 1,
$
the basis elements of \(\TP_5(G)\) map to basis elements of \(\TP_5(G_{\rm Tot})\);
consequently it suffices to trivialize the corresponding anomaly indices themselves,
without constructing an explicit basis for the finite-group theories.

\item We will then apply the appropriate 
 symmetry-extension trivialization method \cite{Wang2017locWWW1705.06728},
 making 
a nonperturbative global anomaly 
 in $G$
 becomes anomaly-free in an appropriate
 $G_{\rm Tot}$ 
 via an appropriate group extension in terms of the following short exact sequence
\bea \label{eq:}
1 \to K \to G_{\rm Tot} \overset{r}{\longrightarrow} 
G 
\to
1.
\eea
Namely,
a nonperturbative global anomaly index 
\bea
\nu_G \in \TP_d(G)
\eea 
in the Freed-Hopkins version
\cite{1604.06527} of cobordism group TP
(see \Sec{sec:Notations})
becomes the trivial anomaly class 
\bea \label{eq:trivial-anomaly-class}
(r^*\nu)_{G_{\rm Tot}} =0 \in \TP_d(G_{\rm Tot})
\eea
for the cobordism group TP of the pulled back 
$G_{\rm Tot}$.
Here $r$ is the reduction map from $G_{\rm Tot} \overset{r}{\longrightarrow} 
G$,
then the $r^*$ with a $*$ denote the 
pullback.
According to  
\cite{Wang2017locWWW1705.06728}, this provides 
a 
(3+1)d anomalous $G$-symmetric  
 $K$-gauge topological order construction
 whose low-energy theory is a (3+1)d $K$-gauge TQFT, which is designed to carry the original
 't Hooft anomaly index
  in $G$, namely the anomaly index $\nu_G \in \TP_d(G)$.
  
 Here the mathematical question is that given a $G$,
how to find suitable $K$ and $G_{\rm Tot}$ to make the condition \eq{eq:trivial-anomaly-class} works out? In principle, it is not an easy question, but we can work out examples by
examples, see \Sec{sec:SEexamples}.

In fact, 
\Sec{sec:SEexamples}'s
examples are largely motivated by the physics properties of the Standard Model (SM) of particle physics
and Beyond the Standard Model (BSM) in \Sec{sec:SManomalies}.

\end{enumerate}

\newpage
\section{$({\bf B \pm L})$ Mixed Gauge-Gravitational Anomalies of the Standard Model}\label{sec:SManomalies}

Now we compute the anomaly index of the Standard Model (SM) involving the mixed gauge-gravitational anomaly of the discrete $({\bf B \pm L})$ internal symmetry and the gravitational background.
SM 
is a 4d chiral gauge theory with Yang-Mills spin-1 gauge fields of
the  Lie algebra 
\bea \label{eq:SMLieAlgebra}
\cG_{\rm SM} \equiv su(3) \times  su(2) \times u(1)_{\tilde Y}
\eea
coupling to $N_f=3$ families of 15 or 16 Weyl fermions (spin-$\frac{1}{2}$ Weyl spinor 
is in the ${\bf 2}_L^\C$ representation {of} the spacetime symmetry Spin(1,3),
written as a left-handed 15- or 16-plet $\psi_L$)
in the following $\cG_{\rm SM}$ representation
\begin{multline}
    \label{eq:SMrep}
({\psi_L})_{\rm I} =
( \bar{d}_R \oplus {l}_L  \oplus q_L  \oplus \bar{u}_R \oplus   \bar{e}_R  
)_{\rm I}
\oplus
n_{\nu_{{\rm I},R}} {\bar{\nu}_{{\rm I},R}}
\\
\sim 
\big((\overline{\bf 3},{\bf 1})_{2} \oplus ({\bf 1},{\bf 2})_{-3}  
\oplus
({\bf 3},{\bf 2})_{1} \oplus (\overline{\bf 3},{\bf 1})_{-4} \oplus ({\bf 1},{\bf 1})_{6} \big)_{\rm I}
\oplus n_{\nu_{{\rm I},R}} {({\bf 1},{\bf 1})_{0}}
\end{multline} 
for each family. 
Hereafter the family index is denoted by {symbols in roman font} ${\rm I}=1,2,3$; 
with ${\psi_L}_1$ for $u,d,e$ type,
${\psi_L}_2$ for $c,s,\mu$ type,
and 
${\psi_L}_3$ for $t,b,\tau$ type of quarks and leptons.
Both of the left-handed particles, 
$q_L$ and $l_L$, are the weak force SU(2) doublets, for quarks and leptons respectively.
The right-handed anti-particles,
up quark $\bar{u}_R$, down quark $\bar{d}_R$, neutrino ${\bar{\nu}_{R}}$, and electron $\bar{e}_R$ are the  weak force SU(2) singlets.
There is also a spin-0 Higgs scalar $\phi$ in $({\bf 1},{\bf 2})_{3}$ but which does not contribute to fermionic anomaly that we concern here. 
We use ${\rm I}=1,2,3$ for $n_{\nu_{e,R}}, n_{\nu_{\mu,R}}, n_{\nu_{\tau,R}} \in \{ 0, 1\}$
to label either the absence or presence of electron $e$, muon $\mu$, or tauon $\tau$ types of sterile neutrinos (i.e., ``right-handed'' neutrinos sterile to $\cG_{\rm SM}$ gauge forces).
Below we consider $N_f$ families 
(typically $N_f=3$) of fermions (including quarks and leptons),
and sterile neutrinos of the total number $n_{\nu_{R}} \equiv \sum_{\rm I} n_{\nu_{{\rm I},R}}$    
{which can be equal, smaller, or larger than 3 (here ${\rm I}=1,2,3,\dots$ for $e,\mu,\tau,\dots$ type of neutrinos)}.

\begin{table}[!h]
\begin{tabular}{|c |  c  | c | c |  c |  c | c  | c|  c | c |  }
\hline
& 
$\bar{d}_R$ & $l_L$ & $q_L$ & $\bar{u}_R$
& $\bar{e}_R= e_L^+$ & $\bar{\nu}_R$ &$\phi_H$ & $\TP_5$ & 
{\parbox{2.5cm}{ 
\vspace{2pt}
$\bar{\nu}_R$
anom index
\vspace{2pt}}}\\
\hline\rule{0pt}{10pt}
${\SU(3)}$ & $\overline{\mathbf{3}}$ & $\mathbf{1}$ & ${{\mathbf{3}}}$ & $\overline{\mathbf{3}}$ & $\mathbf{1}$ & $\mathbf{1}$ & $\mathbf{1}$ & &\\
${\SU(2)}$ & $\mathbf{1}$ & $\mathbf{2}$ & $\mathbf{2}$  & $\mathbf{1}$ & $\mathbf{1}$ & $\mathbf{1}$ & $\mathbf{2}$ & &\\
$\U(1)_{Y}$ & 1/3 & $-1/2$ & 1/6 & $-2/3$ & 1 & 0 & ${1}/{2}$ & &\\
$\U(1)_{\tilde Y }$ & 2 & $-3$ & 1 & $-4$ & 6 & 0 & $3$ & &\\
$\U(1)_{\rm{EM}}$ & 1/3 & 0 or $-1$ & 2/3 or $-1/3$ & $-2/3$ & 1 & 0 & 0 & &\\[1mm]
 \hline\rule{0pt}{10pt}
$\U(1)_{{ \mathbf{B}-  \mathbf{L}}}
=\U(1) ^\rF$ & $-1/3$ & $-1$ & $1/3$ & $-1/3$ & 1 & 1 & 0 & $\Z^2$ &\\
$\U(1)_{{{\bf Q}} - {N_c}{\bf L}}
=\U(1) ^\rF$ & $-1$ & $-3$ & 1 & $-1$ & 3 & 3 & 0 & $\Z^2$ &\\
$\U(1)_{X} =\U(1) ^\rF$ 
& $-3$ & $-3$ & 1 & 1 & 1 & 5 & $-2$ & $\Z^2$ &\\
$\Z_{5,X}$ & 2 & 2 & 1 & 1 & 1 & 0 & $-2$ & $\Z_5^2$ &\\
$\Z_{4,X}= \Z_4^\rF$ & 1 & 1 & 1 & 1 & 1 & 1 & $-2$ & \cpurple{$\Z_{16}$} & \cpurple{1}\\
$\Z_{8,X}= \Z_8^\rF$ & 5 & 5 & 1 & 1 & 1 & 5 & $-2$ & \cpurple{$\Z_{32} \oplus \Z_2$} & \cpurple{$(-7,1)$}\\
%
%
{\parbox{3.8cm}{ 
\vspace{2pt}
$\Z_{ 2N_f=6, {{\bf B}} + {\bf L}}
=  \Z_6^\rF$ \\
for $N_f=3$;\\
or $\Z_2^\rF, \Z_4^\rF$\\
for $N_f=1,2$\\
(broken from 
$\U(1)_{{{\bf B}} + {\bf L}}$).
\vspace{2pt}}}
& $-1/3$ & $1$ & $1/3$ & $-1/3$ & $-1$ & $-1$ & 0
&
{\parbox{1.8cm}{ 
\vspace{2pt}
\cpurple{$\Z_9$}\\
(or \cpurple{0} or \cpurple{$\Z_{16}$})
\vspace{2pt}}}
&
{\parbox{1.8cm}{ 
\vspace{2pt}
\cpurple{$-1$}\\
(or \cpurple{0} or \cpurple{$-1$})
\vspace{2pt}}}
\\
{\parbox{3.8cm}{ 
\vspace{2pt}
$\Z_{ 2N_cN_f=18, {{\bf Q}} + {N_c}{\bf L}}
= \Z_{18}^\rF$ \\
for $N_f=3$;\\
or $\Z_6^\rF, \Z_{12}^\rF$\\
for $N_f=1,2$\\
(broken from 
$\U(1)_{{{\bf Q}} + {N_c}{\bf L}}$).
\vspace{2pt}}}
& $-1$ & $3$ & 1 & $-1$ & $-3$ & $-3$ & 0
&
{\parbox{1.8cm}{ 
\vspace{2pt}
\cpurple{$\Z_{27} \oplus \Z_3$}\\
(or \cpurple{$\Z_9$} or \cpurple{$\Z_{16} \oplus \Z_9$})
\vspace{2pt}}} &
{\parbox{2.cm}{ 
\vspace{2pt}
\cpurple{$(0,1)$ vs $(9,2)$}\\
(or \cpurple{$0$} or \cpurple{$(1,0)$})
\vspace{2pt}}}
\\
$\Z_{2}^\rF$ & 1 & 1 & 1 & 1 & 1 & 1 & 0 & 0 &\\
\hline
\end{tabular}
\caption{Follow the convention in \cite{Putrov:2023jqi2302.14862}, we show the representations of quarks and leptons in terms of  
Weyl fermions 
in various internal symmetry groups.
Each fermion is a spin-$\frac{1}{2}$ Weyl spinor 
${\bf 2}_L$ representation {of} the spacetime symmetry group Spin(1,3).
Each fermion is written as a left-handed particle $\psi_L$ or a right-handed anti-particle $\ii \sigma_2 \psi_R^*$.
The groups above the horizontal line are the gauge groups in the SM energy scale.
The groups below the horizontal line are potential global symmetries at the  SM energy scale, but they might be secretly dynamically gauged at higher-energy if they are anomaly-free.
Here the anomaly index of $\bar{\nu}_R$
has two different bases choices so we have two different anomaly indices \cpurple{$(0,1)$ vs $(9,2)$}
in \cpurple{\Table{table-SpinxZn} vs \Table{table-Spin-Zn}}.
But for three $\bar{\nu}_R$ in the SM, we have the same physics because
\cpurple{$3(0 \mod 27,1 \mod 3)=3(9 \mod 27, 2 \mod 3)
=(0 \mod 27, 0 \mod 3)$.}
}
\label{table:SM}
\end{table}

In \Table{table:SM},
we list down the representation of quarks and leptons in the $({\bf B \pm L})$ and other related symmetries motivated by
\eq{eq:SpinB-LB+L} and \eq{eq:SpinQ-NcLQ+NcL}.
In addition,
a discrete baryon {\bf B} minus lepton {\bf L} symmetry linear combined with a properly  quantized electroweak hypercharge $\tilde{Y}$,
known as $X \equiv 5({\bf B} - {\bf L})-\frac{2}{3} \tilde{Y}
= \frac{5}{N_c}({{\bf Q}- N_c {\bf L}})-\frac{2}{3} \tilde{Y}$ is identified by 
Wilczek-Zee \cite{Wilczek1979hcZee,WilczekZeePLB1979}, which can be continuously  preserved as $\U(1)_X$ in the SM. 
But the $\U(1)_X$ symmetry has the same 't Hooft anomaly captured by \eq{SM-U1-iTFT-2}.

This specific combination as an order four finite cyclic group $\Z_{4, X \equiv 5({\bf B} - {\bf L})-\frac{2}{3} \tilde{Y}}$ is also pointed out \cite{Wilczek1979hcZee,WilczekZeePLB1979}.
The $\Z_{4, X}$ symmetry in the SM is exactly the $\Z_{4}^{\rm F}$ symmetry
discussed in our main text. We can also consider
$\Z_{8, X}$
as $\Z_{8}^{\rm F}$
symmetry.

In \Table{table:SM},
we also list down the (3+1)d fermionic anomaly classification of the given symmetry,
by the TP cobordism group $\TP_5$.
We then list down the 
anomaly index of the 
$\bar{\nu}_R$ 
within $\TP_5$ on the last column.

In \Table{table:SM-2}, 
we extend the \Table{table:SM} further, for various $G={\Spin \times_{\Z_2^\rF} G^\rF}$,
we list down the $\bar{\nu}_R$ charge under $G^\rF$,
we include 
TP cobordism group $\TP_5^G$,
$\bar{\nu}_R$ anomaly index,
$N_f$-family SM's anomaly
index $\upnu$ for the SM missing $N_f$ $\bar{\nu}_R$.
We also comment on whether the anomaly can be canceled by 
anomalous (3+1)d topological orders,
and what would be their (minimal) gauge group $K$ for the
anomalous (3+1)d $K$-gauge theory.

\begin{table}[!h]
\begin{tabular}{|c |  c  | c | c |  c |   c |  c | }
\hline
$G^\rF$ & 
 $\bar{\nu}_R$ & $\TP_5^{\Spin \times_{\Z_2^\rF} G^\rF}$ & 
{\parbox{2.cm}{ 
\vspace{2pt}
$\bar{\nu}_R$\\
anomaly \\
index $\upnu$
\vspace{1.pt}}} &
{\parbox{2.cm}{ 
\vspace{2pt}
$N_f$-family  SM\\
anomaly \\
index $\upnu$
\vspace{2pt}}} 
&
{\parbox{2.cm}{ 
\vspace{2pt}
 anomalous\\
4d \\
topological order
\vspace{2pt}}} 
&
{\parbox{2.cm}{ 
\vspace{2pt}
anomalous\\
4d $K$-gauge\\
theory
\vspace{2pt}}} 
\\
\hline
 \hline\rule{0pt}{10pt}
$\Z_{4,X}= \Z_4^\rF$ &   1 &   \cpurple{$\Z_{16}$} & \cpurple{1} & \cpurple{$-3$}
&
{\parbox{2.cm}{ 
\vspace{2pt}
$\upnu$ even: symm TQFT. \\
$\upnu$ odd: symm fracton.\\
\vspace{2pt}}} 
&
{\parbox{2.cm}{ 
\vspace{2pt}
$\upnu$ even: \\
$\Z_4$-gauge TQFT. \\
$\upnu$ odd:
No (Symmetry-extension\\
impossible).\\
\vspace{2pt}}}
\\
\hline
$\Z_{8,X}= \Z_8^\rF$ &   5 &  \cpurple{$\Z_{32} \oplus \Z_2$} & \cpurple{$(-7,1)$}\ & \cpurple{$(21,1)$} &
{\parbox{2.cm}{ 
\vspace{2pt}
$ \upnu_1 = 0 \; {\rm mod}\; 4$  :  symm TQFT. \\
$ \upnu_1 = 1 \; {\rm mod}\; 2$ : no symm TQFT.\\
$ \upnu_1 = 2 \; {\rm mod}\; 4$ : symm  fracton (?).\\
$ \upnu_2 = 1 \; {\rm mod}\; 2$ : symm TQFT.\\
\vspace{2pt}}}
&
{\parbox{2.cm}{ 
\vspace{2pt}
$ \upnu_1 = 0 \; {\rm mod}\; 4$  :  $\Z_4$-gauge TQFT. \\
$ \upnu_1 = 1 \; {\rm mod}\; 2$ : No.\\
$ \upnu_1 = 2 \; {\rm mod}\; 4$ : No.\\
$ \upnu_2 = 1 \; {\rm mod}\; 2$ : $\Z_4$-gauge TQFT. \\
\vspace{2pt}}}
\\
\hline
{\parbox{3.8cm}{ 
\vspace{2pt}
$\Z_{ 2N_f=2, {{\bf B}} + {\bf L}}
=  \Z_2^\rF$ \\
for $N_f=1$.\\
$\Z_{ 2N_f=4, {{\bf B}} + {\bf L}}
=  \Z_4^\rF$ \\
for $N_f=2$.\\
$\Z_{ 2N_f=6, {{\bf B}} + {\bf L}}
=  \Z_6^\rF$ \\
for $N_f=3$.\\
(broken from 
$\U(1)_{{{\bf B}} + {\bf L}}$).
\vspace{2pt}}}
&  $-1$  
&
{\parbox{1.cm}{ 
\vspace{2pt}
 \cpurple{0}. \\
 \cpurple{$\Z_{16}$}.\\
\cpurple{$\Z_9$}.\\
\vspace{2pt}}}
&
{\parbox{1.cm}{ 
\vspace{2pt}
 \cpurple{0}. \\
 \cpurple{$-1$}.\\
\cpurple{$-1$}.\\
\vspace{2pt}}} &
{\parbox{1.cm}{ 
\vspace{2pt}
 \cpurple{0}. \\
 \cpurple{$2$}.\\
\cpurple{$3$}.\\
\vspace{2pt}}}
&
{\parbox{2.cm}{ 
\vspace{2pt}
For $\Z_4^\rF$,\\
$ \upnu = 0 \; {\rm mod}\; 2$  :  symm TQFT. \\
For $\Z_6^\rF$,\\
$ \upnu = 0 \; {\rm mod}\; 3$ :   symm GC TQFT.
\vspace{2pt}}}
&
{\parbox{2.cm}{ 
\vspace{2pt}
For $\Z_4^\rF$,\\
$ \upnu = 0 \; {\rm mod}\; 2$  :  $\Z_4$-gauge TQFT. \\
For $\Z_6^\rF$,\\
$ \upnu = 0 \; {\rm mod}\; 3$ :   symm GC TQFT.
\vspace{2pt}}}
\\
\hline
{\parbox{3.8cm}{ 
\vspace{2pt}
$\Z_{ 2N_cN_f=6, {{\bf Q}} + {N_c}{\bf L}}
= \Z_{6}^\rF$ \\
for $N_f=1$.\\
$\Z_{ 2N_cN_f=12, {{\bf Q}} + {N_c}{\bf L}}
= \Z_{12}^\rF$ \\
for $N_f=2$.\\
$\Z_{ 2N_cN_f=18, {{\bf Q}} + {N_c}{\bf L}}
= \Z_{18}^\rF$ \\
for $N_f=3$.\\
(broken from 
$\U(1)_{{{\bf Q}} + {N_c}{\bf L}}$).
\vspace{2pt}}}
&   $-3$   &
{\parbox{1.8cm}{ 
\vspace{2pt}
\cpurple{$\Z_9$}. \\
\cpurple{$\Z_{16} \oplus \Z_9$}.\\ 
\cpurple{$\Z_{27} \oplus \Z_3$}.
\vspace{2pt}}} &
{\parbox{1.cm}{ 
\vspace{2pt}
\cpurple{$0$}.\\
 \cpurple{$(1,0)$}.\\
\cpurple{$(0,1)$}.
\vspace{2pt}}} &
{\parbox{1.cm}{ 
\vspace{2pt}
\cpurple{$0$}.\\
 \cpurple{$(-2,0)$}.\\
\cpurple{$(0,-3)$}\\
\cpurple{$=(0,0)$}.
\vspace{2pt}}}
&
{\parbox{2.cm}{ 
\vspace{2pt}
For $\Z_{12}^\rF$,\\
$ \upnu = 0 \; {\rm mod}\; 2$  :  symm TQFT. \\
For $\Z_{18}^\rF$,\\
$ \upnu = (0,0)$ :   No anomaly.
\vspace{2pt}}}
&
{\parbox{2.cm}{ 
\vspace{2pt}
For $\Z_{12}^\rF$,\\
$ \upnu = 0 \; {\rm mod}\; 2$  :  $\Z_4$-gauge TQFT. \\
For $\Z_{18}^\rF$,\\
$ \upnu = (0,0)$ :   No anomaly.
\vspace{2pt}}}
\\
\hline
\end{tabular}
\caption{For various $G={\Spin \times_{\Z_2^\rF} G^\rF}$,
we list down the $\bar{\nu}_R$ charge under $G^\rF$,
we include 
TP cobordism group $\TP_5^G$ anomaly classification,
$\bar{\nu}_R$ anomaly index,
$N_f$-family SM's anomaly
index $\upnu$ for the SM missing $N_f$ $\bar{\nu}_R$.
We also comment on whether the anomaly can be canceled by 
anomalous (3+1)d topological orders,
and what would be their (minimal) gauge group $K$ for the
anomalous (3+1)d $K$-gauge theory.}
\label{table:SM-2}
\end{table}

\newpage

Previously in \Table{table:SM} and \Table{table:SM-2}, we only list down the coefficients of the $\TP_5$ cobordism group classification. 
Here, in the following, we further enumerate, case by case, a choice of the bases
of the $\TP_5$ cobordism group generators, in terms of some 5d
invertible topological field theory 
(iTFT) partition function complex valued in
U(1).

\begin{enumerate}
\item 
\cpurple{$\TP_5(\Spin^c)=\Z  \oplus \Z$}:
 \cpurple{$(1,0)$}
 for $\bar{\nu}_R$ in \Table{table:SM}.

{\bf $\U(1)_{{\bf B}- {\bf L}}$ or $\U(1)_{{\bf Q}- N_c {\bf L}}$ symmetry and anomaly}:

A continuous baryon {\bf B} minus lepton {\bf L} number symmetry, $\U(1)_{{\bf B}- {\bf L}}$ is preserved within the SM.
More precisely, to have a properly quantized charge, it is better to normalize $\U(1)_{{\bf B}- {\bf L}}$ 
by a factor of $N_c$ as
the quark {\bf Q} number minus $N_c$ lepton {\bf L} number symmetry,
$\U(1)_{{\bf Q}- N_c {\bf L}}$, where the color number is $N_c=3$ in the SM.\footnote{Namely,
when we mention {$\U(1)_{{\bf B}- {\bf L}}$ symmetry and anomaly}, we really mean
$\U(1)_{{\bf Q}- N_c {\bf L}}$ symmetry and anomaly.}
But the $\U(1)_{{\bf Q}- N_c {\bf L}}$ symmetry has a 
't Hooft anomaly in 4d spacetime captured by 
a 5d invertible topological field theory (iTFT) in one extra dimension
with the following invertible U(1) functional:
\begin{equation}\label{SM-U1-iTFT-1}
   \exp(\ii  S_5)
\equiv \exp\Bigg[\ii   \int_{M^5} (-N_f+n_{\nu_R}) \,A \wedge \left(N_c^3 \frac{1}{6} \dd A \wedge  \dd A +N_c\frac{1}{24}  \frac{1}{8 \pi^2}   \Tr[ R \wedge R] \right) \Bigg], 
\end{equation}

where $A$ is the U(1) gauge field connection (locally a 1-form) and $R$ is the spacetime curvature locally a 2-form.
In terms of the relation between \eq{SM-U1-iTFT-1} and the perturbative Feynman diagram as triangle cubic term diagrams in \Fig{fig:Feynman}: 

\begin{figure}[!h]
    \centering
\includegraphics[width=.25\columnwidth]{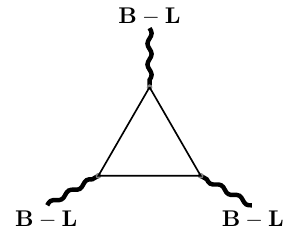}
\includegraphics[width=.25\columnwidth]{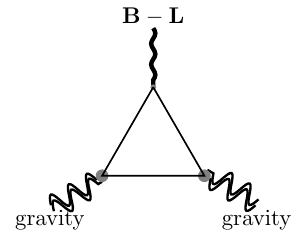}
    \caption{The Standard Model has $\U(1)^3$ anomaly and $\U(1)$-gravity$^2$ anomaly
    with the $\U(1) \equiv \U(1)_{ {\bf L}}$.
    However, because $\U(1)_{ {\bf L}}$
    is only a classical symmetry not a quantum symmetry under the Adler-Bell-Jackiw anomaly with the dynamical SM gauge group, we should consider instead the quantum symmetry
    $\U(1) \equiv \U(1)_{{\bf B}- {\bf L}}$ or more precisely as $\U(1) \equiv \U(1)_{{\bf Q}- N_c {\bf L}}$, captured by two kinds of one-loop triangle Feynman diagrams shown here.
    The anomaly coefficient is given by $(-N_f+n_{\nu_R})$, the integer-value difference between the family number $N_f=3$ and 
    the total number $n_{\nu_R}$ of types of right-handed neutrinos.}
    \label{fig:Feynman}
\end{figure}

the $A \wedge \dd A \wedge  \dd A$ matches with the $\U(1)^3$ anomaly
and the $A \wedge  \Tr[ R \wedge R]$ matches with the $\U(1)$-gravity$^2$ anomaly.
\Eq{SM-U1-iTFT-1} can be written as in terms of Chern and Pontryagin characteristic classes:
\begin{equation}
 \label{SM-U1-iTFT-2}
\exp(\ii  S_5)
\equiv \exp\Bigg[\ii   \int_{M^5} (-N_f+n_{\nu_R}) \,A_{{\bf Q}-N_c {\bf L}}\left(N_c^3 \frac{c_1(\U(1)_{{{\bf Q}-N_c {\bf L}}})^2}{6}-N_c\frac{p_1(TM)}{24}\right) \Bigg] .
\end{equation}
{Here the $n$-th Chern class is denoted as $c_n$, locally $c_1 \coloneqq \int_{M^2} \frac{1}{2 \pi} \dd A$;  
the first Pontrygain class is $p_1$, locally $p_1  \coloneqq \int_{M^4} - \frac{1}{8 \pi^2}   \Tr[ R \wedge R]$. To obtain a nontrivial characteristic class, it requires taking care of the transition functions between local patches glued together to do a global integral on the whole manifold.}

\item  
\cpurple{$\TP_5(\Spin\times_{\Z_2^\rF}\Z_4^\rF)=\Z_{16}$}: \cpurple{$1$}
 for $\bar{\nu}_R$ in \Table{table:SM}.

{\bf $\Z_4^\rF = \Z_{4, X \equiv 5({\bf B} - {\bf L})-\frac{2}{3} \tilde{Y}}$  
symmetry and anomaly}:

Recently, in [\onlinecite{1808.00009, WW2019fxh1910.14668,
JW2006.16996, JW2008.06499, JW2012.15860,
WangWanYou2112.14765, WangWanYou2204.08393, Putrov:2023jqi2302.14862}],
it was pointed out that the (3+1)d SM suffers from a mod 16 class of this mixed $\Z_{4, X}$-gauge-gravity nonperturbative global anomaly captured by a (4+1)d fermionic invertible topological field theory
(namely, the low-energy field theory of the (4+1)d FSPT that we studied)
\bea  \label{SM-Z16-iTFT} 
\exp(\ii S_5^{(\upnu_{})})  
\equiv 
\exp\Bigg[\ii  \upnu\,
     \frac{2\pi }{16}  \eta_{4{\dd}}(\text{PD}( A_{{\Z_{2}}} )) \big\vert_{M^5}\Bigg]
=
\exp\Bigg[\ii  (-N_f+n_{\nu_R})\,
     \frac{2\pi }{16}  \eta_{4{\dd}}(\text{PD}( A_{{\Z_{2}}} )) \big\vert_{M^5}\Bigg].
\eea
The background gauge field $A_{{\Z_{2}}}\in \H^1(M^5,\Z_2)$ is 
obtained by the quotient map $\Z_{2}\equiv {\Z_{4,X}}/{\Z_2^\rF}$ from 
the ${\Spin \times_{\Z_2^\rF} {\Z_{4,X}}}$-structure on the 5d spacetime manifold $M^5$.
The 4d Atiyah-Patodi-Singer eta invariant $\eta_{4{\dd}}$ 
is the $\Z_{16}$ class of topological invariant of time-reversal symmetric topological superconductor (with a time-reversal generator $\rm{T}$ whose
$\rm{T}^2=(-1)^{\rm F}$).
The $\eta_{4{\dd}}$ is evaluated at the 4d submanifold Poincar\'e dual (PD) to the $A_{{\Z_{2,X}}}$ in the 5d bulk.

Let us call the generic anomaly index $\upnu$ 
(here in \eq{SM-Z16-iTFT}, $\upnu = -N_f+n_{\nu_R}$).
Although it is not precise mathematically, based on the chart of the Adams spectral sequence, we can rewrite the 5d iTFT partition function \eq{SM-Z16-iTFT} on a 5d manifold $M^5$ into
\begin{multline} \label{eq:5dSPT-4dTQFT}
\exp(\ii  S_5^{(\upnu_{})})
\equiv
\exp(\frac{2\pi \ii}{16} \cdot\upnu \cdot \eta(\text{PD}(A_{{\Z_2}})) \bigg\rvert_{M^5})\\
= 
\exp(\frac{2\pi \ii}{16} \cdot\upnu \cdot  \Big( 8 \cdot\frac{p_1(TM)}{48}(\text{PD}(A_{{\Z_2}})) + 4 \cdot\text{Arf} (\text{PD}((A_{{\Z_2}})^3) ) 
+ 2 \cdot{\tilde{\eta}} (\text{PD}((A_{{\Z_2}})^4) )   + (A_{{\Z_2}})^5 \Big) \bigg\rvert_{M^5})\\
= 
\exp(\frac{2\pi \ii}{16} \cdot\upnu \cdot  \Big( 8 \cdot\frac{\sigma}{16}(\text{PD}(A_{{\Z_2}})) + 4 \cdot\text{Arf} (\text{PD}((A_{{\Z_2}})^3) ) 
+ 2 \cdot{\tilde{\eta}} (\text{PD}((A_{{\Z_2}})^4) )   + (A_{{\Z_2}})^5 \Big) \bigg\rvert_{M^5}),
\end{multline}
for a generic ${\upnu=-N_{\text{generation}}} \in \Z_{16}$.\\
$\bullet$ The $p_1(TM)$ is the first Pontryagin class of spacetime tangent bundle $TM$ of the manifold $M$.
Via the Hirzebruch signature theorem, we have 
$\frac{1}{3}\int_{\Sigma^4} p_1(TM) =\sigma(\Sigma^4) =\sigma$ ${=\frac{1}{8 \pi^2}\int \Tr(R(\omega) \wedge R(\omega))}$
on a 4-manifold $\Sigma^4$, where $\sigma$ is the signature of $\Sigma^4$ {while $\omega$ is the 1-connection of tangent bundle and
$R(\omega)$ is the Riemann curvature 2-form of $\omega$}.\\
So in \eq{eq:5dSPT-4dTQFT}, we evaluate the $\frac{1}{3}\int_{\Sigma^4} p_1(TM) =\sigma$ on the Poincar\'e dual (PD) $\Sigma^4$ manifold of
 the $(A_{{\Z_2}})$ cohomology class within the $M^5$. If $\Sigma^4$ is a spin manifold, then $\sigma$ is a multiple of 16, hence $\frac{p_1(TM)}{48}(\text{PD}(A_{{\Z_2}}))$ in \eq{eq:5dSPT-4dTQFT} is well-defined.
\\
{$\bullet$ The $\tilde{\eta}$ is a mod 2 index of 1d Dirac operator as a cobordism invariant of $\Omega_1^{\Spin}=\Z_2$. 
{The 1d manifold generator of $\tilde{\eta}$ is a circle $S^1$ with a periodic boundary condition (i.e., Ramond) for the fermion.} Since a 1d manifold is always a spin manifold, ${\tilde{\eta}} (\text{PD}((A_{{\Z_2}})^4) )$ in \eq{eq:5dSPT-4dTQFT} is well-defined.\\
$\bullet$  The Arf invariant \cite{Arf1941} is a mod 2 cobordism invariant of $\Omega_2^{\Spin}=\Z_2$,
 whose quantum matter realization is the 1+1d Kitaev fermionic chain \cite{Kitaev2001chain0010440} {whose each open end hosts a 0+1d Majorana zero mode}. If the Poincar\'e dual (PD) of
 the $(A_{{\Z_2}})^3$ cohomology class within the $M^5$ is a spin manifold, then the Arf invariant $\text{Arf} (\text{PD}((A_{{\Z_2}})^3) )$ in \eq{eq:5dSPT-4dTQFT} is well-defined.\\
$\bullet$ The $(A_{{\Z_2}})^5$ is a mod 2 class purely bosonic topological invariant, which corresponds to a 5d bosonic SPT phase given by
 the group cohomology class data $\H^5(\B\Z_2,\U(1))=\Z_2$, which is also one of the $\Z_2$ generators in $\Omega_5^{\SO}(\B\Z_2)$.
 }
 
\item \cpurple{$\TP_5(\Spin\times_{\Z_2^\rF}\Z_8^\rF)=\Z_{32} \oplus \Z_2$}: \cpurple{$(-7,1)$} for $\bar{\nu}_R$ in \Table{table:SM}.

{\bf $\Z_8^\rF = \Z_{8, X \equiv 5({\bf B} - {\bf L})-\frac{2}{3} \tilde{Y}}$  
symmetry and anomaly}:

The generator of $(1,0)\in \Z_{32} \oplus \Z_2$ is
\bea\label{eq:nu1}
\exp(\frac{2\pi\ii}{32}\cdot\upnu_1\cdot(8\cdot \frac{p_1(TM)}{48}(\text{PD}(A_{\Z_4}))+\mathfrak{P}(\beta_{(4,4)}A_{\Z_4}))).
\eea

The generator of $(0,1)\in \Z_{32} \oplus \Z_2$ is
\bea\label{eq:nu2}
\exp(\pi\ii\cdot\upnu_2\cdot \text{Arf}(\text{PD}(A_{\Z_4}\beta_{(4,4)}A_{\Z_4}))).
\eea
These generators are read from the chart of the Adams spectral sequence.
However, they are not precise mathematically.

The background gauge field $A_{{\Z_{4}}}\in \H^1(M^5,\Z_4)$ is 
obtained by the quotient map $\Z_{4}\equiv {\Z_{8,X}}/{\Z_2^\rF}$ from 
the ${\Spin \times_{\Z_2^\rF} {\Z_{8,X}}}$-structure on the 5d spacetime manifold $M^5$.

$\bullet$ $\beta_{(4,4)}:\H^1(-,\Z_4)\to \H^2(-,\Z_4)$ is the Bockstein homomorphism. \\
$\bullet$ $\mathfrak{P}:\H^2(-,\Z_4)\to \H^5(-,\Z_8)$ is the Postnikov square.\\
$\bullet$ The $p_1(TM)$ is the first Pontryagin class of spacetime tangent bundle $TM$ of the manifold $M$.
Via the Hirzebruch signature theorem, we have 
$\frac{1}{3}\int_{\Sigma^4} p_1(TM) =\sigma(\Sigma^4) =\sigma$ ${=\frac{1}{8 \pi^2}\int \Tr(R(\omega) \wedge R(\omega))}$
on a 4-manifold $\Sigma^4$, where $\sigma$ is the signature of $\Sigma^4$ {while $\omega$ is the 1-connection of tangent bundle and
$R(\omega)$ is the Riemann curvature 2-form of $\omega$}.\\
So in \eq{eq:nu1}, we evaluate the $\frac{1}{3}\int_{\Sigma^4} p_1(TM) =\sigma$ on the Poincar\'e dual (PD) $\Sigma^4$ manifold of
 the $(A_{{\Z_4}})$ cohomology class within the $M^5$. If $\Sigma^4$ is a spin manifold, then $\sigma$ is a multiple of 16, hence $\frac{p_1(TM)}{48}(\text{PD}(A_{{\Z_4}}))$ in \eq{eq:nu1} is well-defined.\\
 $\bullet$  The Arf invariant \cite{Arf1941} is a mod 2 cobordism invariant of $\Omega_2^{\Spin}=\Z_2$,
 whose quantum matter realization is the 1+1d Kitaev fermionic chain \cite{Kitaev2001chain0010440} {whose each open end hosts a 0+1d Majorana zero mode}. If the Poincar\'e dual (PD) of
 the $A_{\Z_4}\beta_{(4,4)}A_{\Z_4}$ cohomology class within the $M^5$ is a spin manifold, then the Arf invariant $\text{Arf} (\text{PD}(A_{\Z_4}\beta_{(4,4)}A_{\Z_4}) )$ in \eq{eq:nu2} is well-defined.

\item \cpurple{$\TP_5(\Spin\times\Z_3)=\Z_{9}$}: \cpurple{$-1$} for $\bar{\nu}_R$ in \Table{table:SM}.

By \Table{table-SpinxZn}, the anomaly index of a charge $q=1$ Weyl fermion with symmetry $\Spin\times\Z_3$ is $1\in\TP_5(\Spin\times\Z_3)=\Z_9$. Now we give the explicit basis of $\TP_5(\Spin\times\Z_3)=\Z_9$.

We start with a perturbative local anomaly of U(1) charge $q=1$ left-handed Weyl fermion 
in $\Spin \times \U(1)$ to
derive the
the nonperturbative global anomaly 
of $\Z_{3, {\bf B + L}}$ charge $q=1$ Weyl fermion 
in 
$\Spin \times \Z_{3, {\bf B + L}}$.

The perturbative local anomaly of U(1) charge $q=1$ left-handed Weyl fermion 
of $\Spin \times \U(1)$ symmetry
in 3+1d or 4d
is captured by a 5d invertible field theory (iTFT) with the anomaly index $k=1$:
\begin{equation}
   \exp( \ii k  \int_{M^5} A \frac{c_1^2}{6}-A \frac{p_1}{24}).
\end{equation} 
Now we redefine the U(1) gauge field $A$ as a $\Z_3$ gauge field $ A_{\Z_3} \in\H^1(\B\Z_3,\Z_3)= \Z_3$
with the following replacement:
\bea
A &\mapsto& \frac{2 \pi}{3}   A_{\Z_3}.\cr 
c_1 = \frac{\dd A}{2 \pi} &\mapsto& 
\frac{\dd   A_{\Z_3}}{3} \equiv \beta_{(3,3) } A_{\Z_3}. 
\eea
The $\beta_{(n,m)}: \H^*(-,\Z_m) \mapsto 
\H^{*+1}(-,\Z_n) $ 
is the Bockstein 
 homomorphism associated with the extension
 $\Z_n \stackrel{\cdot m}{\to} \Z_{nm} \to \Z_m$. Thus we get the 5d topological invariant of the $\Spin \times \Z_3$ as:
\bea\label{eq:Spin-Z3}
   &&\exp \big( \ii 2 \pi k  \int_{M^5} 
   ( \frac{1}{18} 
   {A_{\Z_3}} (\beta_{(3,3)} {A_{\Z_3}}) (\beta_{(3,3)} {A_{\Z_3}})
   -\frac{1}{3 \cdot 24} {A_{\Z_3}} p_1 ) \big) \cr
   &=&\exp \big( \ii \frac{2 \pi}{9} k  \int_{M^5} ( \frac{1}{2} 
   {A_{\Z_3}} (\beta_{(3,3)} {A_{\Z_3}}) (\beta_{(3,3)} {A_{\Z_3}})
   -\frac{1}{8} {A_{\Z_3}} p_1 ) \big)\cr
   &=&\exp \big( \ii \frac{2 \pi}{9} k  \int_{M^5} ( -4
   {A_{\Z_3}} (\beta_{(3,3)} {A_{\Z_3}}) (\beta_{(3,3)} {A_{\Z_3}})
   +3\cdot {} \frac{A_{\Z_3}p_1}{3} ) \big).
\eea 
Here $\beta_{(3,3)}:\H^1(-,\Z_3)\to \H^2(-,\Z_3)$ is the Bockstein homomorphism.
Here ${} \frac{A_{\Z_3}p_1}{3}$ is a mod 3 class that involves Pontryagin class because $A_{\Z_3}p_1=0\mod3$ \cite{tomonaga1964mod,tomonaga1965pontryagin},
while ${A_{\Z_3}} (\beta_{(3,3)} {A_{\Z_3}}) (\beta_{(3,3)} {A_{\Z_3}})$
is a mod 3 class.

In \eqref{eq:Spin-Z3}, the first quality rewrites the coefficients $\frac{1}{18}=\frac{1}{9}\cdot\frac{1}{2}$ and $-\frac{1}{3\cdot 24}=\frac{1}{9}\cdot(-\frac{1}{8})$ since the anomaly of 4d Weyl fermion with symmetry $\Spin\times\Z_3$ contains only 3-torsion \cite{Kapustin1406.7329, 2016arXiv160406527F, 1808.00009, 1808.02881, GuoJW1812.11959, 2506.19710}, we can regard $2$ and $8$ as invertible in $\Z_9$.\\
The second equality uses the fact that $1=-8\mod9$ to obtain $\frac{1}{2}=-4\mod9$ and $-\frac{1}{8}=1\mod9$ and uses the fact that $A_{\Z_3}p_1=0\mod3$ \cite{tomonaga1964mod,tomonaga1965pontryagin} to rewrite $A_{\Z_3}p_1=3\cdot \frac{A_{\Z_3}p_1}{3}$.

%

Thus the 4d fermionic anomaly 
has the anomaly index $k \in \Z_{9}$, agreeing with the bordism group classification by
$\Omega_5^{{\rm Spin} \times \Z_{3}}=\Z_9$
\cite{Kapustin1406.7329, 2016arXiv160406527F, 1808.00009, 1808.02881, GuoJW1812.11959, 2506.19710}.

Therefore, the generator of $1\in\Z_{9}=\TP_5(\Spin\times\Z_3)$ is 
\bea
\exp \big( \ii \frac{2 \pi}{9}   \int_{M^5} ( -4
   {A_{\Z_3}} (\beta_{(3,3)} {A_{\Z_3}}) (\beta_{(3,3)} {A_{\Z_3}})
   +3\cdot {} \frac{A_{\Z_3}p_1}{3} ) \big).
\eea

\item  \cpurple{$\TP_5(\Spin\times\Z_9)=\TP_5(\Spin\times_{\Z_2^{\rF}}\Z_{18}^{\rF})=\Z_{27} \oplus \Z_3$}: \cpurple{$(0,1)$ vs $(9,2)$} for $\bar{\nu}_R$ in \Table{table:SM}.

By \Table{table-Spin-Zn}, the anomaly indices of a charge $q=1$ Weyl fermion with symmetry $\Spin\times_{\Z_2^{\rF}}\Z_{18}$ are $(9,2)\in\TP_5(\Spin\times_{\Z_2^{\rF}}\Z_{18})=\Z_{27} \oplus \Z_3$.
Now we give the explicit basis of $(1,0)\in \TP_5(\Spin\times_{\Z_2^{\rF}}\Z_{18})=\Z_{27} \oplus \Z_3$.

We start with a perturbative local anomaly of U(1) charge $q=1$ left-handed Weyl fermion 
in $\Spin^c$ to
derive the
the nonperturbative global anomaly 
of $\Z_{18, {\bf B + L}}^{\rF}$ charge $q=1$ Weyl fermion 
in 
$\Spin \times_{\Z_2^{\rF}} \Z_{18, {\bf B + L}}^{\rF}$.

We compare the $\Spin^c$ gauge field and the U(1) gauge field.
{For $\Spin^c$, 
 the $\U(1) \supset {\Z_2^\rF}$ contains the fermion parity as a normal subgroup.\\
$\bullet$ For the original $\U(1)$ with $c_1(\U(1))$, 
the gauge bundle constraint is $w_2(TM)= 2 c_1 \mod 2$.
In the original $\U(1)$, fermions have odd charges under $\U(1)$,
while bosons have even charges under $\U(1)$.
We call the original U(1) gauge field $A$,
then $c_1=\frac{\dd A}{2 \pi} \in \frac{1}{2}\Z$.\\
$\bullet$ For the new $\U(1)'=\frac{\U(1)}{\Z_2^\rF}$ with $c_1(\U(1)')$,
the gauge bundle constraint is $w_2(TM)= c_1' = 2 c_1 \mod 2$.
We call the new $\U(1)'$ gauge field $A'$,
then $c_1'=\frac{\dd A'}{2 \pi}=\frac{\dd (2A)}{2 \pi} = 2 c_1\in 2\frac{1}{2}\Z = \Z$.\\
$\bullet$ To explain why $A' = 2 A$ or $ c_1' = 2 c_1$, we look at the Wilson line operator
$\text{$\exp(\ii q' \oint A')$ and $\exp(\ii q \oint A)$.}$
The original $\U(1)$ has charge transformation $\exp(\ii q \theta)$ with $\theta \in [0, 2 \pi)$,
while the new $\U(1)'$ has charge transformation $\exp(\ii q' \theta')$ with $\theta' \in [0, 2 \pi)$.
But the $\U(1)'=\frac{\U(1)}{\Z_2^\rF}$, so the $\theta=\pi$ in the old $\U(1)$ 
is identified as $\theta'=2\pi$ as a trivial zero
in the new $\U(1)'$.
In the original $\U(1)$, the $q \in \Z$ to be compatible with $\theta \in [0, 2 \pi)$.
In the new $\U(1)'$, the original $q$ is still allowed to have $2\Z$ to be compatible with $\theta \in [0, \pi)$;
but the new $q'=\frac{1}{2} q \in \Z$
and the new $\theta'= 2 \theta  \in [0, 2 \pi)$ are scaled accordingly.
Since the new $q'=\frac{1}{2} q \in \Z$, we show the new $A'=2 A$.}

The perturbative local anomaly of charge $q=1$ left-handed Weyl fermion 
of $\Spin^c$ symmetry
in 3+1d or 4d
is captured by a 5d invertible field theory (iTFT) with the anomaly index $k'=1$:
\begin{equation}
   \exp( \ii k'  \int_{M^5} A' \frac{(2c_1)^2}{48}-A' \frac{p_1}{48}).
\end{equation} 
Now we redefine the U(1) gauge field $A'$ as a $\Z_9$ gauge field $ A'_{\Z_9} \in\H^1(\B\Z_9,\Z_9)= \Z_9$
with the following replacement:
\bea
A' &\mapsto& \frac{2 \pi}{9}   A'_{\Z_9}.\cr 
2c_1=c_1' = \frac{\dd A'}{2 \pi} &\mapsto& 
\frac{\dd   A'_{\Z_9}}{9} \equiv \beta_{(9,9) } A'_{\Z_9}. 
\eea
The $\beta_{(n,m)}: \H^*(-,\Z_m) \mapsto 
\H^{*+1}(-,\Z_n) $ 
is the Bockstein 
 homomorphism associated with the extension
 $\Z_n \stackrel{\cdot m}{\to} \Z_{nm} \to \Z_m$. Thus we get the 5d topological invariant of the $\Spin \times_{\Z_2^{\rF}}\Z_{18}$ as:
\bea\label{eq:Spin-Z18-Z2}
   &&\exp \big( \ii 2 \pi k'  \int_{M^5} 
   ( \frac{1}{9\cdot48} 
   {A'_{\Z_9}} (\beta_{(9,9)} {A'_{\Z_9}}) (\beta_{(9,9)} {A'_{\Z_9}})
   -\frac{1}{9 \cdot 48} {A'_{\Z_9}} p_1 ) \big) \cr
   &=&\exp \big( \ii \frac{2 \pi}{27} k'  \int_{M^5} ( \frac{1}{16} 
   {A'_{\Z_9}} (\beta_{(9,9)} {A'_{\Z_9}}) (\beta_{(9,9)} {A'_{\Z_9}})
   -\frac{1}{16} {A'_{\Z_9}} p_1 ) \big)\cr
   &=&\exp \big( \ii \frac{2 \pi}{27} k'  \int_{M^5} ( 
   -5{A'_{\Z_9}} (\beta_{(9,9)} {A'_{\Z_9}}) (\beta_{(9,9)} {A'_{\Z_9}})
   +5\cdot 3\cdot {} \frac{A'_{\Z_9}p_1}{3} ) \big).
\eea  
Here $\beta_{(9,9)}:\H^1(-,\Z_9)\to \H^2(-,\Z_9)$ is the Bockstein homomorphism.
Here ${} \frac{A'_{\Z_9}p_1}{3}$ is a mod 9 class that involves Pontryagin class because $A'_{\Z_9}p_1=0\mod3$ \cite{tomonaga1964mod,tomonaga1965pontryagin},
while ${A'_{\Z_9}} (\beta_{(9,9)} {A'_{\Z_9}}) (\beta_{(9,9)} {A'_{\Z_9}})$
is a mod 9 class.

In \eqref{eq:Spin-Z18-Z2}, the first equality rewrites the coefficients $\frac{1}{9\cdot 48}=\frac{1}{27}\cdot\frac{1}{16}$ since the anomaly of 4d Weyl fermion with symmetry $\Spin\times_{\Z_2^{\rF}}\Z_{18}$ contains only 3-torsion \cite{Kapustin1406.7329, 2016arXiv160406527F, 1808.00009, 1808.02881, GuoJW1812.11959, 2506.19710}, we can regard $16$ as invertible in $\Z_{27}$.\\
The second equality uses the fact that $1=-80\mod27$ to obtain $\frac{1}{16}=-5\mod27$ and uses the fact that $A'_{\Z_9}p_1=0\mod3$ \cite{tomonaga1964mod,tomonaga1965pontryagin} to rewrite $A'_{\Z_9}p_1=3\cdot \frac{A'_{\Z_9}p_1}{3}$.

Therefore, the generator of $(1,0)\in\Z_{27} \oplus \Z_3=\TP_5(\Spin\times_{\Z_2^{\rF}}\Z_{18})$ is
\bea
\exp \big( \ii \frac{2 \pi}{27}  \int_{M^5} ( 
   -5{A'_{\Z_9}} (\beta_{(9,9)} {A'_{\Z_9}}) (\beta_{(9,9)} {A'_{\Z_9}})
   +5\cdot 3\cdot {} \frac{A'_{\Z_9}p_1}{3} ) \big).
\eea



\end{enumerate}

\newpage
\section{Examples of Symmetry Extension
 to Trivialize Fermionic Anomalies:\\
 (3+1)d Anomalous $G$-Symmetric  
 $K$-Gauge Topological Order}\label{sec:SEexamples}
 
 In this section, we study some examples of symmetry extension
 to trivialize fermionic anomalies by replacing the Weyl fermion theory with a (3+1)d $G$-symmetric anomalous 
 $K$-gauge topological order.

\subsection{Multiple Weyl fermions with 
$G=\Spin\times\Z_4$-symmetric charges $q \in \Z_4$
under extension:\\
$1 \to K=\Z_2\to G_{\rm Tot}=\Spin\times\Z_8 \to G=\Spin\times\Z_4 \to 1$
}

We consider multiple left-handed Weyl fermions $\psi_q$ under the $G=\Spin\times\Z_4$ symmetry, with charge 
$q=0,1,2,3 \in \Z_4$ respectively 
for each fermion that potentially has a
$G$-anomaly. We will list down the anomaly index in $G$,
and the possible symmetry extension to trivialize the anomaly 
by pulling back $G$ to $G_{\rm Tot}$.

For the representation
$R=\rho_{q}\in RU(\Z_4)$, the unitary representation ring of the group $\Z_4$,
we have $q=0,1,2,3 \in \Z_4$.

For $G=\Spin\times\Z_4$, 
for each Weyl fermion $\psi_{q}$ with a charge $q \in \Z_4$,
the $\Z_4$ is generated by 
\bea
\U_{\theta}:
\psi_q\to \e^{\ii q\theta}\psi_q ,
\eea
where $\theta=\frac{2\pi}{4}k$, $k\in\Z_4$. 
The anomaly indices $(\nu_1'(n,q),\nu_2'(n,q))$ for $\TP_5(\Spin\times\Z_n)=\Z_{a_n}\times\Z_{b_n}$ are given in \eq{eq:nu'} \cite{2506.19710},
where $a_n$ and $b_n$ are defined in \eq{eq:Spin-Zn-TP5}.
In particular, the anomaly index of a charge $q$ left-handed Weyl fermion $\psi_q$ has only the first index,
$(\nu_1'(4,q),\nu_2'(4,q))=(\nu_1'(4,q),0)$,
with
\bea
\nu_1'(n=4,q) = \frac{a_n(n^2+3n+2)q^3}{6n} {\mid}_{n=4}=
q^3 \mod4 \text{ in } \TP_5(\Spin\times\Z_4)=\Z_4, 
\eea
where we use $n=4$, $a_4=4$, and $5q^3=q^3\mod4$.
In short,
the anomaly index of a charge $q=0,1,2,3$ 
left-handed Weyl fermion $\psi_q$ is 
\bea \label{eq:table-SpinxZ4}
\begin{tabular}{|c |  c  | c | c |  c | }
\hline
$q$ & 0 & 1 & 2 & 3\\
\hline
\hline
$\nu_1'(n=4, q) \mod 4$ & 0 & 1 & 0 & $3=-1$\\
\hline
\end{tabular}
\eea

We consider the following symmetry extension to trivialize the anomaly:
\bea
1\to K=\Z_2\to G_{\rm Tot}=\Spin\times\Z_8\to G=\Spin\times\Z_4\to1.
\eea

In $G_{\rm Tot}=\Spin\times\Z_8$, 
the $\Z_8$ is generated by 
\bea
\U'_{\theta'}:\psi_{q'}\to \e^{\ii q'\theta'}\psi_{q'}
\eea
where $\theta'=\frac{2\pi}{8}k'$, $k'\in\Z_8$.
Since $k=k'\mod4$ 
due to the quotient relation $G_{\rm Tot}/\Z_2=G$,
therefore
we can impose $q'=2q$
for the pull back
so that $\U_{\theta}=\U'_{\theta'}$. 

The normal subgroup $K=\Z_2$ is generated by $\U'_{\theta',k'=4}$.

After the fermions are pulled back from the symmetry 
$G$ to the symmetry $G_{\rm Tot}$,
we have a new charge
$q'=2q$. 
Namely, the induced representation is $R'=\rho_{q'=2q}
\in RU(\Z_8)$, the unitary representation ring of the group $\Z_8$.

The anomaly indices for $\Spin\times\Z_8$ symmetry,
with the classification 
$\TP_5(\Spin\times\Z_8)=\Z_8\times\Z_2$,
are
\bea\label{eq:nu1'n=8q'}
\nu_1'(n=8,q') &=& \frac{a_n(n^2+3n+2)q'^3}{6n} {\mid}_{n=8}=
- q'^3 \mod 8 \text{ in }  \Z_8 \text{ of } 
\text{ in } \TP_5(\Spin\times\Z_8)=\Z_8 \oplus \Z_2. \\\label{eq:nu2'n=8q'}
\nu_2'(n=8,q') &=& \frac{2b_n(q'-q'^3)}{n} {\mid}_{n=8}=
\frac{q'-q'^3}{2} \mod 2 \text{ in } \Z_2 \text{ of } 
\TP_5(\Spin\times\Z_8)=\Z_8 \oplus \Z_2. 
\eea

In short,
the anomaly index of a charge $q' \in \Z_8$ 
left-handed Weyl fermion $\psi_{q'}$ is 
\bea \label{eq:table-SpinxZ8}
\begin{tabular}{|c |  c  | c | c |  c |  c  | c | c |  c | }
\hline
$q'$ & 0 & 1 & 2 & 3 & 4 & 5 & 6 & 7 \\
\hline
\hline
$\nu_1'(n=8,q')\mod 8$ & 0 & $-1=7$ & 0 & $-3=5$ &  0 & 3  & 0  & 1\\
\hline
$\nu_2'(n=8,q')\mod 2$ & 0 & 0 & 1 & 0 & 0 & 0 & 1 & 0\\
\hline
\end{tabular}
\eea
It is physically more intuitive to normalize by a $(-1)$ sign (e.g., by flipping the basis)
to the $\nu_1'(n=8,q'=1)=1$, so 
the charge $q'=1$ Weyl fermion becomes the 
$\nu_1'=1$ anomaly index generator, so 
\eq{eq:table-SpinxZ8} becomes
:
\bea \label{eq:table-SpinxZ8-2}
\begin{tabular}{|c |  c  | c | c |  c |  c  | c | c |  c | }
\hline
$q'$ & 0 & 1 & 2 & 3 & 4 & 5 & 6 & 7 \\
\hline
\hline
$\nu_1'(n=8,q')\mod 8$ & 0 & $1$ & 0 & $3$ &  0 & $5$  & 0  & $7$\\
\hline
\end{tabular}
\eea
By pulling back $\psi_q$ in $G$ to $\psi_{q'=2q}$ in $G_{\rm Tot}$,
combining the info in
 \eq{eq:table-SpinxZ4}, \eqref{eq:table-SpinxZ8}, and \eqref{eq:table-SpinxZ8-2}:
\begin{enumerate}
    \item Odd number of $\psi_{q=1}$ (or $\psi_{q=3}$) Weyl fermions' $G$-anomaly cannot be trivialized by $K=\Z_2$ extension, as odd number of $\psi_{q'=2}$ (or $\psi_{q'=6}$) Weyl fermions in $G_{\rm Tot}$ still has $G_{\rm Tot}$-anomaly.
     \item Even number of $\psi_{q=1}$ (or $\psi_{q=3}$) Weyl fermions' $G$-anomaly can be trivialized by $K=\Z_2$ extension, as even number of $\psi_{q'=2}$ (or $\psi_{q'=6}$) Weyl fermions in $G_{\rm Tot}$ is $G_{\rm Tot}$-anomaly free.
      \item Equal numbers of $\psi_{q=1}$ and $\psi_{q=3}$ Weyl fermions are anomaly-free,
      thus anomaly is already trivial without symmetry extension.
\end{enumerate}

\subsection{Multiple 
Weyl fermions with 
$G=\Spin\times_{\Z_2^{\rF}}\Z_4$-symmetric 
charges $q = 1,3 \in \Z_4$
under extension:\\
$1 \to K=\Z_2\to G_{\rm Tot}=\Spin\times\Z_4 \to G=\Spin\times_{\Z_2^{\rF}}\Z_4 \to 1$}
\label{sec:Gtot=SpinxZ4}
 
We consider multiple left-handed Weyl fermions $\psi_q$ under the $G=\Spin\times_{\Z_2^{\rF}}\Z_4$ symmetry, with charge $q = 1,3 \in \Z_4$ respectively for each fermion that potentially has a $G$-anomaly. We will list down the anomaly index in $G$,
and the possible symmetry extension via $K=\Z_2$ to trivialize the anomaly 
by pulling back $G$ to $G_{\rm Tot}$.

For the representation $\tilde{R}=\tilde{\rho}_{q}\in RU^o(\Z_4)$, the additive subgroup of $RU(\Z_4)$ generated by the 1-dimensional representations $\tilde{\rho}_q$ of the group $\Z_4$ with odd charges $q$, we have $q=1,3 \in \Z_4$.

For $G=\Spin\times_{\Z_2^{\rF}}\Z_4$, for each Weyl fermion $\psi_q$ with a charge $q=1,3 \in \Z_4$, the $\Z_4$ is generated by 
\bea \label{eq:Uq}
\U_{\theta}:\psi_q\to \e^{\ii q\theta}\psi_q,
\eea
where $\theta=\frac{2\pi}{4}k$, $k\in\Z_4$. 
For even $m$, the anomaly indices $(\tilde{\nu}_1'(m,q),\tilde{\nu}_2'(m,q))$ for $\TP_5(\Spin\times_{\Z_2^{\rF}}\Z_{2m})=\Z_{\tilde{a}_m}\times\Z_{\tilde{b}_m}$ are given in \eq{eq:tildenu'} \cite{2506.19710}, where $\tilde{a}_m$ and $\tilde{b}_m$ are defined in \eq{eq:Spin-Z2m-TP5}.
In particular, the anomaly index of a charge $q$ left-handed Weyl fermion $\psi_q$ has only the first index, $(\tilde{\nu}_1'(2,q),\tilde{\nu}_2'(2,q))=(\tilde{\nu}_1'(2,q),0)$ with
\bea \label{eq:nu1m=2q}
\tilde{\nu}_1'(m=2,q)=\tilde{a}_m\frac{(2m^2+m+1)q^3-(m+3)q}{48m}|_{m=2}=\frac{11q^3-5q}{6}\mod16 \text{ in }\TP_5(\Spin\times_{\Z_2^{\rF}}\Z_4)=\Z_{16},
\eea
where we use $m=2$ and $\tilde{a}_2=16$.
In short,
the anomaly index of a charge $q=1,3$ 
left-handed Weyl fermion $\psi_q$ is 
\bea \label{eq:table-SpinxZ2Z4}
\begin{tabular}{| c | c |  c | }
\hline
$q$ &  1  & 3\\
\hline
\hline
$\tilde{\nu}_1'(m=2,q) \mod 16$ &$1$&$47=-1$\\
\hline
\end{tabular}
\eea

We consider the following symmetry extension to trivialize the anomaly:
\bea
1\to K=\Z_2\to G_{\rm Tot}=\Spin\times\Z_4\to G=\Spin\times_{\Z_2^{\rF}}\Z_4\to1.
\eea

In $G_{\rm Tot}=\Spin\times\Z_4$, the $\Z_4$ is generated by 
\bea
\U'_{\theta'}:\psi_{q'}\to \e^{\ii q'\theta'}\psi_{q'}
\eea
where $\theta'=\frac{2\pi}{4}k'$, $k'\in\Z_4$.
Since $k=k'\mod2$ due to the quotient relation $G_{\rm Tot}/\Z_2=G$, therefore we can impose $q'=q$ for the pull back so that $\U_{\theta}=\U'_{\theta'}$.

The normal subgroup $K=\Z_2$ is generated by 
 the diagonal element $((-1)^{\rF},\U'_{\theta',k'=2})$ in $G_{\rm Tot}=\Spin\times\Z_4$, where $(-1)^{\rF}$
 is the fermion parity such that for any fermion $\psi$,
 $$
 (-1)^{\rF} \psi = -  \psi. 
$$
After the fermions are pulled back from the symmetry 
$G$ to the symmetry $G_{\rm Tot}$,
we have a new charge $q'=q$. Namely, the induced representation is $\tilde{R}'=\rho_{q'=q}\in RU(\Z_4)$, the unitary representation ring of the group $\Z_4$.

The anomaly indices $(\nu_1'(n,q),\nu_2'(n,q))$ for $\TP_5(\Spin\times\Z_n)=\Z_{a_n}\times\Z_{b_n}$ are given in \eq{eq:nu'} \cite{2506.19710},
where $a_n$ and $b_n$ are defined in \eq{eq:Spin-Zn-TP5}.
In particular, the anomaly index of a charge $q'$ left-handed Weyl fermion $\psi_{q'}$ has only the first index,
$(\nu_1'(4,q'),\nu_2'(4,q'))=(\nu_1'(4,q'),0)$,
with
\bea
\nu_1'(n=4,q') = \frac{a_n(n^2+3n+2)q'^3}{6n} {\mid}_{n=4}=
q'^3 \mod4 \text{ in } \TP_5(\Spin\times\Z_4)=\Z_4, 
\eea
where we use $n=4$, $a_4=4$, and $5q'^3=q'^3\mod4$.
In short,
the anomaly index of a charge $q'\in\Z_4$ 
left-handed Weyl fermion $\psi_{q'}$ is 
\bea \label{eq:table-SpinxZ4'}
\begin{tabular}{|c |  c   |  c | c   |  c | }
\hline
$q'$ &0& 1 &2&  3\\
\hline
\hline
$\nu_1'(n=4, q') \mod 4$ & 0& 1 &0& $3=-1$\\
\hline
\end{tabular}
\eea

By pulling back $\psi_q$ in $G$ to $\psi_{q'=q}$ in $G_{\rm Tot}$, combining the info in \eq{eq:table-SpinxZ2Z4} and \eq{eq:table-SpinxZ4'}:
\begin{enumerate}
  \item $N=1,2,3\mod 4$ copies of $\psi_{q=1}$ (or $\psi_{q=3}$) Weyl fermions' $G$-anomaly cannot be trivialized by $K=\Z_2$ extension, as $N=1,2,3\mod 4$ copies of $\psi_{q'=1}$ (or $\psi_{q'=3}$) Weyl fermions in $G_{\rm Tot}$  still has $G_{\rm Tot}$-anomaly.
     \item $N=0\mod4$ copies of $\psi_{q=1}$ (or $\psi_{q=3}$) Weyl fermions' $G$-anomaly can be trivialized by $K=\Z_2$ extension, as $N=0\mod4$ copies of $\psi_{q'=1}$ (or $\psi_{q'=3}$) Weyl fermions in $G_{\rm Tot}$ is $G_{\rm Tot}$-anomaly free.
      \item Equal numbers of $\psi_{q=1}$ and $\psi_{q=3}$ Weyl fermions are anomaly-free,
      thus anomaly is already trivial without symmetry extension.
\end{enumerate}

\subsection{Multiple 
Weyl fermions with 
$G=\Spin\times_{\Z_2^{\rF}}\Z_4$-symmetric 
charges $q = 1,3 \in \Z_4$
under extension:\\
$1 \to K=\Z_4\to G_{\rm Tot}=\Spin\times\Z_8 \to G=\Spin\times_{\Z_2^{\rF}}\Z_4 \to 1$}

We consider multiple left-handed Weyl fermions $\psi_q$ under the $G=\Spin\times_{\Z_2^{\rF}}\Z_4$ symmetry, with charge $q = 1,3 \in \Z_4$ respectively for each fermion that potentially has a $G$-anomaly. We will list down the anomaly index in $G$,
and the possible symmetry extension via $K=\Z_4$ to trivialize the anomaly 
by pulling back $G$ to $G_{\rm Tot}$.

The initial setup is essentially the same as in Section \ref{sec:Gtot=SpinxZ4}, we also follow 
\eq{eq:Uq}, \eqref{eq:nu1m=2q}, \eqref{eq:table-SpinxZ2Z4}.

Now we consider the following new 
symmetry extension to trivialize the anomaly:
\bea
1\to K=\Z_4\to G_{\rm Tot}=\Spin\times\Z_8\to G=\Spin\times_{\Z_2^{\rF}}\Z_4\to 1.
\eea

In $G_{\rm Tot}=\Spin\times\Z_8$, the $\Z_8$ is generated by 
\bea
\U'_{\theta'}:
\psi\to \e^{\ii q'\theta'}\psi
\eea
where $\theta'=\frac{2\pi}{8}k'$, $k'\in\Z_8$.
Since $k=k'\mod4$ due to the quotient relation $G_{\rm Tot}/\Z_4=G$, therefore we can impose $q'=2q$ for the pull back so that $\U_{\theta}=\U'_{\theta'}$. 

The normal subgroup $K=\Z_4$ is generated by the diagonal element 
\bea
\text{$((-1)^{\rF},\U'_{\theta',k'=2})$ in $G_{\rm Tot}=\Spin\times\Z_8$,}
\eea
where $(-1)^{\rF}$ corresponds to the fermion parity $\Z_2^{\rF}$ normal subgroup of the Spin so $\Spin/\Z_2^{\rF}=\SO$, while $\U'_{\theta',k'=2}$
generates the $\Z_4$ normal subgroup of $\Z_8$.
It is easy to check that this diagonal element $((-1)^{\rF},\U'_{\theta',k'=2})$ generates an order-4 cyclic group because
$$
((-1)^{\rF},\U'_{\theta',k'=2})^2=(+1, (-1)^{q'}),
\quad
((-1)^{\rF},\U'_{\theta',k'=2})^4= (+1,+1).
$$

After the fermions are pulled back from the symmetry 
$G$ to the symmetry $G_{\rm Tot}$,
we have a new charge $q'=2q$. Namely, the induced representation is $\tilde{R}'=\rho_{q'=2q}\in RU(\Z_8)$, the unitary representation ring of the group $\Z_8$.

The anomaly indices for $\Spin\times\Z_8$ symmetry,
with the classification 
$\TP_5(\Spin\times\Z_8)=\Z_8\times\Z_2$,
are
given in \eq{eq:nu1'n=8q'}, \eqref{eq:nu2'n=8q'}, \eqref{eq:table-SpinxZ8}, and \eqref{eq:table-SpinxZ8-2}.

By pulling back $\psi_q$ in $G$ to $\psi_{q'=2q}$ in $G_{\rm Tot}$,
combining the info in
 \eq{eq:table-SpinxZ2Z4}, \eqref{eq:table-SpinxZ8}, and \eqref{eq:table-SpinxZ8-2}:
 \begin{enumerate}
    \item Odd number of $\psi_{q=1}$ (or $\psi_{q=3}$) Weyl fermions' $G$-anomaly cannot be trivialized by $K=\Z_4$ extension, as odd number of $\psi_{q'=2}$ (or $\psi_{q'=6}$) Weyl fermions in $G_{\rm Tot}$ still has $G_{\rm Tot}$-anomaly.
     \item Even number of $\psi_{q=1}$ (or $\psi_{q=3}$) Weyl fermions' $G$-anomaly can be trivialized by $K=\Z_4$ extension, as even number of $\psi_{q'=2}$ (or $\psi_{q'=6}$) Weyl fermions in $G_{\rm Tot}$ is $G_{\rm Tot}$-anomaly free.
      \item Equal numbers of $\psi_{q=1}$ and $\psi_{q=3}$ Weyl fermions are anomaly-free,
      thus anomaly is already trivial without symmetry extension.
\end{enumerate}

\subsection{Multiple 
Weyl fermions with 
$G=\Spin\times_{\Z_2^{\rF}}\Z_8$-symmetric 
charges $q = 1,3,5,7 \in \Z_8$
under extension:\\
$1 \to K=\Z_4\to G_{\rm Tot}=\Spin\times\Z_{16} \to G=\Spin\times_{\Z_2^{\rF}}\Z_8 \to 1$}

We consider multiple left-handed Weyl fermions $\psi_q$ under the $G=\Spin\times_{\Z_2^{\rF}}\Z_8$ symmetry, with charge $q = 1,3,5,7 \in \Z_8$ respectively for each fermion that potentially has a $G$-anomaly. We will list down the anomaly index in $G$,
and the possible symmetry extension via $K=\Z_4$ to trivialize the anomaly 
by pulling back $G$ to $G_{\rm Tot}$.

For the representation $\tilde{R}=\tilde{\rho}_{q}\in RU^o(\Z_8)$, the additive subgroup of $RU(\Z_8)$ generated by the 1-dimensional representations $\tilde{\rho}_q$ of the group $\Z_8$ with odd charges $q$, we have $q=1,3,5,7 \in \Z_8$.

For $G=\Spin\times_{\Z_2^{\rF}}\Z_8$, for each Weyl fermion $\psi_q$ with a charge $q=1,3,5,7 \in \Z_8$, the $\Z_8$ is generated by 
\bea 
\U_{\theta}:\psi_q\to \e^{\ii q\theta}\psi_q,
\eea
where $\theta=\frac{2\pi}{8}k$, $k\in\Z_8$. 
For even $m$, the anomaly indices $(\tilde{\nu}_1'(m,q),\tilde{\nu}_2'(m,q))$ for $\TP_5(\Spin\times_{\Z_2^{\rF}}\Z_{2m})=\Z_{\tilde{a}_m}\times\Z_{\tilde{b}_m}$ are given in \eq{eq:tildenu'} \cite{2506.19710}, where $\tilde{a}_m$ and $\tilde{b}_m$ are defined in \eq{eq:Spin-Z2m-TP5}.
In particular, the anomaly indices of a charge $q$ left-handed Weyl fermion $\psi_q$ are $(\tilde{\nu}_1'(4,q),\tilde{\nu}_2'(4,q))$ with
\bea 
\tilde{\nu}_1'(m=4,q)=\tilde{a}_m\frac{(2m^2+m+1)q^3-(m+3)q}{48m}|_{m=4}=\frac{37q^3-7q}{6}\mod32 \cr\text{ in }\Z_{32}\text{ of }\TP_5(\Spin\times_{\Z_2^{\rF}}\Z_8)=\Z_{32}\times\Z_2,\\
\tilde{\nu}_2'(m=4,q)=\tilde{b}_m\frac{(m+1)(q^3-q)}{4m}|_{m=4}=\frac{5(q^3-q)}{8}\mod2 \text{ in }\Z_{2}\text{ of }\TP_5(\Spin\times_{\Z_2^{\rF}}\Z_8)=\Z_{32}\times\Z_2
\eea
where we use $m=4$, $\tilde{a}_4=32$, and $\tilde{b}_4=2$.
In short,
the anomaly indices of a charge $q=1,3,5,7$ 
left-handed Weyl fermion $\psi_q$ are 
\bea \label{eq:table-SpinxZ2Z8}
\begin{tabular}{| c | c |  c | c |  c| }
\hline
$q$ &  1  & 3 & 5 & 7\\
\hline
\hline
$\tilde{\nu}_1'(m=4,q) \mod 32$ &$5$&$163=3$&$765=-3$&$2107=-5$\\
\hline
$\tilde{\nu}_2'(m=4,q) \mod 2$ &$0$&$1$&$1$&$0$\\
\hline
\end{tabular}
\eea
It is physically more intuitive to normalize by multiplying the $\tilde{\nu}_1'$ by $13$
to the $\tilde{\nu}_1'(m=4,q=1)=1$, so 
the charge $q=1$ Weyl fermion becomes the 
$\tilde{\nu}_1'=1$ anomaly index generator, so 
\eq{eq:table-SpinxZ2Z8} becomes
:
\bea \label{eq:table-SpinxZ2Z8-2}
\begin{tabular}{|c |  c  | c | c |  c  | }
\hline
$q'$ &  1  & 3 &  5  & 7 \\
\hline
\hline
$\tilde{\nu}_1'(m=4,q) \mod 32$ & $1$ & $7$ & $-7$ & $-1$\\
\hline
\end{tabular}
\eea

We consider the following symmetry extension to trivialize the anomaly:
\bea
1\to K=\Z_4\to G_{\rm Tot}=\Spin\times\Z_{16}\to G=\Spin\times_{\Z_2^{\rF}}\Z_8\to 1.
\eea

In $G_{\rm Tot}=\Spin\times\Z_{16}$, the $\Z_{16}$ is generated by 
\bea
\U'_{\theta'}:
\psi\to \e^{\ii q'\theta'}\psi
\eea
where $\theta'=\frac{2\pi}{16}k'$, $k'\in\Z_{16}$.
Since $k=k'\mod8$ due to the quotient relation $G_{\rm Tot}/\Z_4=G$, therefore we can impose $q'=2q$ for the pull back so that $\U_{\theta}=\U'_{\theta'}$. 

The normal subgroup $K=\Z_4$ is generated by the diagonal element $((-1)^{\rF},\U'_{\theta',k'=4})$ in $G_{\rm Tot}=\Spin\times\Z_{16}$.

After the fermions are pulled back from the symmetry 
$G$ to the symmetry $G_{\rm Tot}$,
we have a new charge $q'=2q$. Namely, the induced representation is $\tilde{R}'=\rho_{q'=2q}\in RU(\Z_{16})$, the unitary representation ring of the group $\Z_{16}$.

The anomaly indices $(\nu_1'(n,q),\nu_2'(n,q))$ for $\TP_5(\Spin\times\Z_n)=\Z_{a_n}\times\Z_{b_n}$ are given in \eq{eq:nu'} \cite{2506.19710},
where $a_n$ and $b_n$ are defined in \eq{eq:Spin-Zn-TP5}.
In particular, the anomaly indices of a charge $q'$ left-handed Weyl fermion $\psi_{q'}$ are
$(\nu_1'(16,q'),\nu_2'(16,q'))$,
with
\bea
\nu_1'(n=16,q') = \frac{a_n(n^2+3n+2)q'^3}{6n} {\mid}_{n=16}=
3q'^3 \mod16 \text{ in } \Z_{16}\text{ of }\TP_5(\Spin\times\Z_{16})=\Z_{16}\times\Z_4, \\
\nu_2'(n=16,q') =\frac{2b_n(q'-q'^3)}{n} {\mid}_{n=16}=\frac{q'-q'^3}{2}\mod4 \text{ in } \Z_{4}\text{ of }\TP_5(\Spin\times\Z_{16})=\Z_{16}\times\Z_4
\eea
where we use $n=16$, $a_{16}=16$, $b_{16}=4$, and $51q'^3=3q'^3\mod4$.
In short,
the anomaly indices of a charge $q'\in\Z_{16}$ 
left-handed Weyl fermion $\psi_{q'}$ is 
\bea \label{eq:table-SpinxZ16}
\begin{tabular}{|c |  c   |  c |c   |  c |c |  c   |  c |c   |  c|c   |  c |c |  c   |  c |c   |  c | }
\hline
$q'$ &0&1& 2 &3&4&5&  6&7&8&9 & 10 &11&12&13& 14&15 \\
\hline
\hline
$\nu_1'(n=16, q') \mod 16$ &0&3& 8& 1& 0&7&8& 5& 0 &11& 8& 9 & 0& 15&8&13 \\
\hline
$\nu_2'(n=16, q') \mod 4$ & 0& 0& 1& 0& 2& 0& 3& 0& 0&0&1&0&2&0&3&0 \\
\hline
\end{tabular}
\eea
It is physically more intuitive to normalize by multiplying the $\nu_1'$ by $11$
to the $\nu_1'(n=16,q'=1)=1$, so 
the charge $q'=1$ Weyl fermion becomes the 
$\nu_1'=1$ anomaly index generator, so 
\eq{eq:table-SpinxZ16} becomes
:
\bea \label{eq:table-SpinxZ16-2}
\begin{tabular}{|c |  c   |  c |c   |  c |c |  c   |  c |c   |  c|c   |  c |c |  c   |  c |c   |  c |    }
\hline
$q'$ &0&1& 2 &3&4&5&  6&7&8&9 & 10 &11&12&13& 14&15  \\
\hline
\hline
$\nu_1'(n=16,q')\mod 16$ &0&1&8&11&0&13&8&7&0&9&8&3&0&5&8&15 \\
\hline
\end{tabular}
\eea
By pulling back $\psi_q$ in $G$ to $\psi_{q'=2q}$ in $G_{\rm Tot}$,
combining the info in
 \eq{eq:table-SpinxZ2Z8}, \eqref{eq:table-SpinxZ2Z8-2}, \eqref{eq:table-SpinxZ16} and \eqref{eq:table-SpinxZ16-2}:

\begin{enumerate}
  \item $N=1,2,3\mod 4$ copies of $\psi_{q=1}$ (or $\psi_{q=3}$ or $\psi_{q=5}$ or or $\psi_{q=7}$) Weyl fermions' $G$-anomaly cannot be trivialized by $K=\Z_4$ extension, as $N=1,2,3\mod 4$ copies of $\psi_{q'=2}$ (or $\psi_{q'=6}$ or $\psi_{q'=10}$ or $\psi_{q'=14}$) Weyl fermions in $G_{\rm Tot}$  still has $G_{\rm Tot}$-anomaly.
     \item $N=0\mod4$ copies of $\psi_{q=1}$ (or $\psi_{q=3}$ or $\psi_{q=5}$ or or $\psi_{q=7}$) Weyl fermions' $G$-anomaly can be trivialized by $K=\Z_4$ extension, as $N=0\mod4$ copies of $\psi_{q'=2}$ (or $\psi_{q'=6}$ or $\psi_{q'=10}$ or $\psi_{q'=14}$) Weyl fermions in $G_{\rm Tot}$ is $G_{\rm Tot}$-anomaly free.
      \item Equal numbers of $\psi_{q=1}$ and $\psi_{q=7}$ (or $\psi_{q=3}$ and $\psi_{q=5}$) Weyl fermions are anomaly-free,
      thus anomaly is already trivial without symmetry extension.
\end{enumerate}

\section{General Statements: Theorems and Proofs}\label{sec:theoremproof}
In this section, we state and prove some general statements
on the symmetry-extension 
$1 \to K \to G_{\rm Tot} \overset{r}{\longrightarrow} G 
\to 1$ 
trivialization by $K$
for fermionic anomalies in $G$ of 
the 2-torsion,
the 3-torsion,
and the $s$-torsion nature. 

We denote
${{\rm Spin}^{\Z_{2^{p+1}  \cdot  3^r \cdot s
}}} \equiv \frac{{\rm Spin} \times {\Z_{2^{p+1}  \cdot  3^r \cdot s
}}}{\Z_2^\rF}
\equiv {{\rm Spin} \times_{\Z_2^\rF} {\Z_{2^{p+1}  \cdot  3^r \cdot s
}}}$.
Here $p \geq 1$, $r \geq 1$, and $2 \nmid s$ and $3 \nmid s$.
The bordism group classifying the 4d  fermionic anomalies with symmetry ${\rm Spin}^{\Z_{2^{p+1}  \cdot  3^r \cdot s
}}$ is \cite{1808.02881}
\bea
&&\Omega_5^{{\rm Spin}^{\Z_{2^{p+1}  \cdot  3^r \cdot s
}}} 
\cong
\Omega_5^{{\rm Spin}^{\Z_{2^{p+1} 
}}}\oplus 
{\Omega}_5^{\rm Spin}(\B\Z_{3^r \cdot s}  )
\cong
\Omega_5^{{\rm Spin}^{\Z_{2^{p+1} 
}}}\oplus \tilde{\Omega}_5^{\rm SO}(\B\Z_{3^r \cdot s}  )
 \cr
&&=
\Z_{2^{p+3}}\oplus \Z_{2^{p-1}}\oplus
\Z_{3^{r+1}}\oplus \Z_{3^{r-1}}\oplus 
\Z_s \oplus 
\Z_s,
\text{ \quad $p \geq 1$, $r \geq 1$, and $2 \nmid s$ and $3 \nmid s$.}
\eea
Here $\tilde{\Omega}_5^{\rm SO}(\B G):=\Omega_5^{\rm SO}(\B G)/\Omega_5^{\rm SO}$ is the reduced bordism group, modding out the $\Omega_5^{\rm SO}=\Omega_5^{\rm SO}(pt)$.

The bordism group classifying the 4d  fermionic anomalies with symmetry ${\rm Spin} \times {\Z_{2^{p}  \cdot  3^r \cdot s
}}$ is \cite{1808.02881}
\bea
&&\Omega_5^{{\rm Spin} \times {\Z_{2^{p}  \cdot  3^r \cdot s
}}} 
\cong
\Omega_5^{{\rm Spin}\times{\Z_{2^{p} 
}}}\oplus \Omega_5^{\Spin}(\B\Z_{3^r \cdot s}  )
\cong
\Omega_5^{{\rm Spin}\times{\Z_{2^{p} 
}}}\oplus \tilde{\Omega}_5^{\rm SO}(\B\Z_{3^r \cdot s}  )
 \cr
&&=
\Z_{2^{p}}\oplus \Z_{2^{p-2}}\oplus
\Z_{3^{r+1}}\oplus \Z_{3^{r-1}}\oplus 
\Z_s \oplus 
\Z_s,
\text{ \quad $p \geq 2$, $r \geq 1$, and $2 \nmid s$ and $3 \nmid s$.}
\eea


As we mentioned in \Sec{sec:Notations}, the TP group classifying the 4d fermionic anomalies with symmetry $\Spin\times\Z_n$ or $\Spin\times_{\Z_2^{\rF}}\Z_{2m}$ is the same as the corresponding bordism group.

Below is a summary of the symmetry-extension trivialization results for charge-$q=1$ Weyl fermion anomalies.

\begin{enumerate}
  \item \textbf{2-torsion.} A symmetry extension can trivialize anomalies for multiples of $2^k$ Weyl fermions (with $k=p$ or $k=p+1$). In particular, either a $K=\mathbb{Z}_4$ (for $k=p$) or a $K=\mathbb{Z}_2$ (for $k=p+1$) extension suffices.
  \item \textbf{3-torsion.} Anomalies for multiples of $3$ Weyl fermions can be trivialized; a $K=\mathbb{Z}_{3^r}$ extension works. These anomalies are of group-cohomology type and therefore admit bosonic analogues.
  \item \textbf{$s$-torsion} (with $2\nmid s$ and $3\nmid s$). For $s$ coprime to $2$ and $3$ (e.g.\ $s=5,7,11,\dots$ or products of such primes), a $K=\mathbb{Z}_s$ extension trivializes the anomaly of a single charge-$q=1$ Weyl fermion. Again, the anomaly must be of group-cohomology type (hence has a bosonic analogue).
\end{enumerate}





\subsection{$2^k$ Weyl fermions with 
$G=\Spin \times_{\Z_2^{\rF}} \Z_{2^{k+1}}$-symmetric 
unit charge
under extension:\\
$1 \to K=\Z_4\to G_{\rm Tot}=\Spin \times \Z_{2^{k+2}}\to G=\Spin \times_{\Z_2^{\rF}} \Z_{2^{k+1}}\to 1$
}
{\bf Theorem 1} In general, the anomaly of $2^k$ Weyl fermions with charge $q=1$ in $\TP_5(\Spin \times_{\Z_2^{\rF}} \Z_{2^{k+1}})$ becomes trivialized by the $\Z_4$ extension in $\TP_5(\Spin \times \Z_{2^{k+2}})$:
 \bea
1\to K=\Z_4\to G_{\rm Tot}=\Spin \times \Z_{2^{k+2}}\to G=\Spin \times_{\Z_2^{\rF}} \Z_{2^{k+1}}\to 1.
\eea

\textit{Proof}. For even $m$, the anomaly indices for $\TP_5(\Spin\times_{\Z_2^{\rF}}\Z_{2m})=\Z_{\tilde{a}_m}\times\Z_{\tilde{b}_m}$ are $\tilde{a}_m\frac{(2m^2+m+1)q^3-(m+3)q}{48m}$ and $\tilde{b}_m\frac{(m+1)(q^3-q)}{4m}$ \cite{2506.19710}.
The anomaly indices for $\TP_5(\Spin\times\Z_n)=\Z_{a_n}\times\Z_{b_n}$ are $\frac{a_n(n^2+3n+2)q^3}{6n}$ and $\frac{2b_n(q-q^3)}{n}$ \cite{2506.19710}.

In particular, the anomaly indices for $\TP_5(\Spin \times_{\Z_2^{\rF}} \Z_{2^{k+1}})=\Z_{2^{k+3}}\times\Z_{2^{k-1}}$ are $\frac{(2^{2k+1}+2^k+1)q^3-(2^k+3)q}{6}\mod 2^{k+3}$ and $\frac{(2^k+1)(q^3-q)}{8}\mod 2^{k-1}$.
The anomaly indices for $\TP_5(\Spin\times\Z_{2^{k+2}})=\Z_{2^{k+2}}\times\Z_{2^k}$ are $\frac{(2^{k+2}+1)(2^{k+2}+2)q^3}{6}\mod 2^{k+2}$ and $\frac{q-q^3}{2}\mod 2^k$. 

For the representation $\tilde{R}=\tilde{\rho}_1^{\oplus2^k}\in RU^o(\Z_{2^{k+1}})$, the charges of Weyl fermions with symmetry $G$ are $q_i=1$ and the charges of Weyl fermions with symmetry $G_{\rm Tot}$ are $q_i'=2$ for $1\le i\le 2^k$. So the induced representation is $\tilde{R}'=\rho_2^{\oplus2^k}\in RU(\Z_{2^{k+2}})$.

The anomaly for $\tilde{R}$ is $2^k\times(\frac{2^{2k}-1}{3},0)\in \Z_{2^{k+3}}\times\Z_{2^{k-1}}$, so it is nontrivial. The anomaly for $\tilde{R}'$ is $2^k\times(\frac{4(2^{k+2}+1)(2^{k+2}+2)}{3},-3)\in \Z_{2^{k+2}}\times\Z_{2^k}$, so it is trivial.
\qed
\subsection{$2^{k+1}$ Weyl fermions with 
$G=\Spin \times_{\Z_2^{\rF}} \Z_{2^{k+1}}$-symmetric 
unit charge  
under extension:\\
$1 \to K=\Z_2\to G_{\rm Tot}=\Spin \times \Z_{2^{k+1}}\to G=\Spin \times_{\Z_2^{\rF}} \Z_{2^{k+1}}\to 1$
}

{\bf Theorem 2} In general, the anomaly of $2^{k+1}$ Weyl fermions with charge $q=1$ in $\TP_5(\Spin \times_{\Z_2^{\rF}} \Z_{2^{k+1}})$ becomes trivialized by the $\Z_2$ extension in $\TP_5(\Spin \times \Z_{2^{k+1}})$:
 \bea
1\to K=\Z_2\to G_{\rm Tot}=\Spin \times \Z_{2^{k+1}}\to G=\Spin \times_{\Z_2^{\rF}} \Z_{2^{k+1}}\to 1.
\eea
\textit{Proof}. For the representation $\tilde{R}=\tilde{\rho}_1^{\oplus2^{k+1}}\in RU^o(\Z_{2^{k+1}})$, the charges of Weyl fermions with symmetry $G$ are $q_i=1$ and the charges of Weyl fermions with symmetry $G_{\rm Tot}$ are $q_i'=1$ for $1\le i\le 2^{k+1}$. So the induced representation is $\tilde{R}'=\rho_1^{\oplus2^{k+1}}\in RU(\Z_{2^{k+1}})$.

Similarly as in the proof of Theorem 1, the anomaly for $\tilde{R}$ is $2^{k+1}\times(\frac{2^{2k}-1}{3},0)\in \Z_{2^{k+3}}\times\Z_{2^{k-1}}$, so it is nontrivial.
Because $\TP_5(\Spin \times \Z_{2^{k+1}})=\Z_{2^{k+1}}\times\Z_{2^{k-1}}$, the order of each element of $\TP_5(\Spin \times \Z_{2^{k+1}})$ divides $2^{k+1}$. Since the anomaly for $\tilde{R}'$ is a multiple of $2^{k+1}$, it is trivial.
\qed

\subsection{$2^k$ Weyl fermions with 
$G=\Spin \times \Z_{2^{k+1}}$-symmetric 
unit charge  
under extension:\\
$1\to K=\Z_2\to G_{\rm Tot}=\Spin \times \Z_{2^{k+2}}\to G=\Spin \times \Z_{2^{k+1}}\to 1$}

{\bf Theorem 3} In general, the anomaly of $2^k$ Weyl fermions with charge $q=1$ in $\TP_5(\Spin \times \Z_{2^{k+1}})$ becomes trivialized by the $\Z_2$ extension in $\TP_5(\Spin \times \Z_{2^{k+2}})$:
 \bea
1\to K=\Z_2\to G_{\rm Tot}=\Spin \times \Z_{2^{k+2}}\to G=\Spin \times \Z_{2^{k+1}}\to 1.
\eea
\textit{Proof}. The anomaly indices for $\TP_5(\Spin\times\Z_n)=\Z_{a_n}\times\Z_{b_n}$ are $\frac{a_n(n^2+3n+2)q^3}{6n}$ and $\frac{2b_n(q-q^3)}{n}$ \cite{2506.19710}.

In particular, the anomaly indices for $\TP_5(\Spin \times \Z_{2^{k+1}})=\Z_{2^{k+1}}\times\Z_{2^{k-1}}$ are $\frac{(2^{k+1}+1)(2^{k+1}+2)q^3}{6}\mod 2^{k+1}$ and $\frac{q-q^3}{2}\mod 2^{k-1}$.
The anomaly indices for $\TP_5(\Spin\times\Z_{2^{k+2}})=\Z_{2^{k+2}}\times\Z_{2^k}$ are $\frac{(2^{k+2}+1)(2^{k+2}+2)q^3}{6}\mod 2^{k+2}$ and $\frac{q-q^3}{2}\mod 2^k$. 

For the representation $R=\rho_1^{\oplus2^k}\in RU(\Z_{2^{k+1}})$, the charges of Weyl fermions with symmetry $G$ are $q_i=1$ and the charges of Weyl fermions with symmetry $G_{\rm Tot}$ are $q_i'=2$ for $1\le i\le 2^k$. So the induced representation is $R'=\rho_2^{\oplus2^k}\in RU(\Z_{2^{k+2}})$.

The anomaly for $R$ is $2^k\times(\frac{(2^{k+1}+1)(2^{k+1}+2)}{6},0)\in \Z_{2^{k+1}}\times\Z_{2^{k-1}}$, so it is nontrivial. The anomaly for $R'$ is $2^k\times(\frac{4(2^{k+2}+1)(2^{k+2}+2)}{3},-3)\in \Z_{2^{k+2}}\times\Z_{2^k}$, so it is trivial.
\qed

\section*{Acknowledgements}

JW thanks Xingyu Ren,
and Matthew Yu for helpful discussions.
JW thanks Meng Cheng and Xinping Yang for the collaboration of
\cite{Cheng:2024awi2411.05786}.
During the completion of this manuscript, in October, we becomes aware that
Debray-Ye-Yu \cite{Debray:2025kfg2510.24834} provides a different approach to solve the similar question:
``How to Construct Anomalous (3+1)d Topological Quantum Field Theories via Symmetry Extension approach?''
ZW is supported by the NSFC Grant No. 12405001.
JW is supported by LIMS and Ben Delo Fellowshop.
JW would like to thank the Isaac Newton Institute for Mathematical Sciences, Cambridge, for support and hospitality during the programme Diving Deeper into Defects: On the Intersection of Field Theory, Quantum Matter, and Mathematics, where work on this paper was undertaken. This work was supported by EPSRC grant EP/Z000580/1.
JW also thanks Simons Foundation Collaboration on Global Categorical Symmetries Annual Meetings in 2024 and 2025, where this work is discussed and performed during the 
meetings.

\bibliography{BSM-TQDM-2511}

\end{document}